\documentclass[sigconf, nonacm]{acmart}

\usepackage{hyperref}
\usepackage[normalem]{ulem} %
\usepackage{xcolor} 
\usepackage{subcaption}
\usepackage{cleveref} %
\usepackage{pifont}
\usepackage{numprint} %
\usepackage{multirow}
\usepackage{makecell} %
\usepackage{wasysym} %

\usepackage{url}

\usepackage{breakurl}

\usepackage{enumitem} %
\npdecimalsign{.}
\npthousandsep{,}

\usepackage[htt]{hyphenat} %
\Crefformat{app}{Appendix #2#1#3}
\Crefformat{table}{Table #2#1#3}

\def\rot{\rotatebox}
\newcommand*{\myalign}[2]{\multicolumn{1}{#1}{#2}}

\newcommand{\empirical}[1]{\setlength{\fboxsep}{1pt}\fbox{#1}}
\renewcommand{\empirical}[1]{#1}

\newcommand{\hide}[1]{}

\newcommand\PartnerA{Cisco} %
\newcommand\PartnerB{InQuest Labs} %

\newcommand\pPDF{clickbait PDF}
\newcommand\PPDF{Clickbait PDF}
\newcommand\cluster{cluster}
\newcommand\Cluster{Cluster}

\newcommand\contribCisco{\numprint{55}} %
\newcommand\contribInquest{\numprint{43}} %
\newcommand\sharedCisInq{\numprint{0.02}} %

\newcommand\nAnalyzedSamples{\numprint{176208}}
\newcommand\nPhashRemoved{\numprint{155535}}
\newcommand\nSamplesAfterPhash{\numprint{20671}}

\newcommand\nSamplesHomogeneousAfterDC{\numprint{18557}}

\newcommand\nDBSCANSamples{\numprint{2114}}
\newcommand\nDBSCANClusters{\numprint{120}}
\newcommand\nDBSCANnoise{\numprint{1135}}
\newcommand\nDBSCANHomCl{\numprint{87}}

\newcommand\nLabels{\numprint{80}}

\newcommand\nMaliciousCampaigns{\numprint{44}}

\newcommand\nUrlsSubmVt{\numprint{19935}}

\newcommand\distinctPageOne{\numprint{157623}}

\newcommand\seoCampaingPercentage{\empirical{\numprint{89}}}

\newcommand\possibleDirectories{\empirical{\numprint{898450}}}
\newcommand\foundIndexPages{\empirical{\numprint{426}}}
\newcommand\totalPdfCrawled{\empirical{\numprint{13012}}}

\newcommand\dirWithUpload{\empirical{\numprint{137}}}

\newcommand\minUrlHeuristic{\empirical{\numprint{11}}}

\newcommand\avgURLrecaptcha{\empirical{\numprint{16}}}
\newcommand\avgURLroblox{\empirical{\numprint{30}}}

\newcommand\exactBing{\empirical{\numprint{3469}}}
\newcommand\exactGoogle{\empirical{\numprint{0}}}
\newcommand\avgExactBing{\empirical{\numprint{59.81}}}
\newcommand\avgExactGoogle{\empirical{\numprint{0}}}
\newcommand\heuristicBing{\empirical{\numprint{6022}}}
\newcommand\heuristicGoogle{\empirical{\numprint{925}}}
\newcommand\avgHeuristicBing{\empirical{\numprint{103.83}}}
\newcommand\avgHeuristicGoogle{\empirical{\numprint{15.95}}}

\newcommand\totalQueries{\empirical{\numprint{47795}}}
\newcommand\exactTotal{\empirical{\numprint{3469}}}
\newcommand\heuristcTotal{\empirical{\numprint{6947}}}

\begin{document}

	\begin{abstract}
	\PPDF s are PDF documents that do not embed malware but trick victims into visiting malicious web pages leading to attacks like password theft or drive-by download. While recent reports indicate a surge of \pPDF s, prior works have largely neglected this new threat, considering PDFs only as accessories of email phishing campaigns. 
	
	This paper investigates the landscape of \pPDF s and presents the first systematic and comprehensive study of this phenomenon. Starting from a real-world dataset, we identify \empirical{\nMaliciousCampaigns{}} \pPDF{} \cluster s via clustering and characterize them by looking at their volumetric, temporal, and visual features. Among these, we identify three large \cluster s covering \empirical{\seoCampaingPercentage\%} of the dataset, exhibiting significantly different volumetric and temporal properties compared to classical email phishing, and relying on web UI elements as visual baits. 
	Finally, we look at the distribution vectors and show that \pPDF s are not only distributed via attachments but also via Search Engine Optimization attacks, placing \pPDF s outside the email distribution ecosystem.
	
	\PPDF s seem to be a lurking threat, not subjected to any form of content-based filtering or detection: AV scoring systems, like VirusTotal, rank them considerably low, creating a blind spot for organizations. While URL blocklists can help to prevent victims from visiting the attack web pages, we observe that they have a limited coverage.
\end{abstract}

\title{From Attachments to SEO: \href{https://www.youtube.com/watch?v=dQw4w9WgXcQ}{\textcolor{blue}{\underline{Click Here}}} to Learn More about \PPDF s!}

\author{Giada Stivala}
\affiliation{%
	\institution{Cispa Helmholtz Center for Information Security}
	\country{}
}
\email{giada.stivala@cispa.de}

\author{Sahar Abdelnabi}
\affiliation{%
	\institution{Cispa Helmholtz Center for Information Security}
	\country{}
}
\email{sahar.abdelnabi@cispa.de}

\author{Andrea Mengascini}
\affiliation{%
	\institution{Cispa Helmholtz Center for Information Security}
	\country{}
}
\email{andrea.mengascini@cispa.de}

\author{Mariano Graziano}
\affiliation{%
	\institution{Cisco Talos}
	\country{}
}
\email{magrazia@cisco.com}

\author{Mario Fritz}
\affiliation{%
	\institution{Cispa Helmholtz Center for Information Security}
	\country{}
}
\email{fritz@cispa.de}

\author{Giancarlo Pellegrino}
\affiliation{%
	\institution{Cispa Helmholtz Center for Information Security}
	\country{}
}
\email{pellegrino@cispa.de}

\maketitle
	
\thispagestyle{plain}
\pagestyle{plain}

	\section{Introduction} %

Phishing emails are one of the major online threats~\cite{verizon_dbir}, where the attacker sends fraudulent emails often attaching PDF files with embedded exploit code or malware~\cite{le2017broad, le2014look, thomas2015framing, simoiu2020targeted}, which compromise victims' computers upon opening the attachments.
Recent reports~\cite{microsoftPhishingAttachments,paloalto_pdftrend} have shown another malicious use of PDF files, which stands out due to a surge in their numbers (estimated in the order of \empirical{five} million files only in 2020~\cite{paloalto_pdftrend}) and to the increased effectiveness of the deceitfulness of visual baits~\cite{microsoftPhishingAttachments,paloalto_pdftrend}. Such PDFs, hereinafter \pPDF s, do not embed any malware or exploit code but only clickbait images which, when clicked, take the victim to an attack webpage stealing passwords, user identities, or compromising victims' computers via drive-by downloads~\cite{microsoftPhishingAttachments,paloalto_pdftrend}.

Although these reports show that large amounts of PDFs lead to attacks on the Web rather than installing malware, the scientific community has largely neglected the threat posed by \pPDF{} files and, to the best of our knowledge, did not investigate the role of PDFs outside classical email phishing attacks.
Prior works have thoroughly explored classical phishing attacks, from empirical measurements of email phishing campaigns' number, volume and temporal dynamics (e.g.,~\cite{simoiu2020targeted}), to studying the duration of phishing attacks (e.g.,~\cite{oest2020sunrise}), including the characteristics of their baits  (e.g.,~\cite{van2019cognitive}), and their effectiveness (e.g.,~\cite{sheng2010falls, dhamija2006phishing}). 
Such works only considered PDFs in the context of email phishing campaigns, however, it is unclear whether \pPDF s are part of them and, if so, to which extent. 
This paper aims to fill this knowledge gap by presenting the first comprehensive study centered on \pPDF{} files.
We study this phenomenon and discuss its evolution, distinctive characteristics, and distribution channels, including their distribution as email attachments.

\paragraph{Our study.} 
We start from a dataset of \nAnalyzedSamples{} PDFs by identifying and clustering PDFs that exhibit meaningful visual similarities. For this analysis, we prioritize content in the first page of the PDF, which, being displayed first to victim users, is most likely to embed an attack bait. 
When the first page also contains a URL, we verify its maliciousness by using a URL analysis service and by manual inspection. Having identified which \cluster s contain PDFs leading to attacks on the Web, we study their temporal and volumetric properties, as well as visual baits and geographical reach.
Finally, the inspection of the structure and visual baits of PDFs leading to Web attacks leads us to hypothesize about two possible distribution vectors, namely email attachments and SEO attacks. We show that \pPDF s analogous to those in our dataset can be found on two search engines, and that online scoring services (i.e., VirusTotal) struggle in clearly separating benign from \pPDF s.

\paragraph{Findings.} Overall, the main finding of our study is providing evidence that PDF files are no longer only ancillary tools of email phishing campaigns. 
Starting from a dataset of \empirical{\nAnalyzedSamples{}} PDF files---collected from Dec. 16th, 2020 to Jun. 23rd, 2021 by two industrial partners---we identified \empirical{\nMaliciousCampaigns{}} out of a total of \empirical{\nLabels{}} \cluster s of \pPDF s whose documents lead to attacks like credential phishing and malware download.
Among \pPDF s, we discovered three \cluster s with significantly different features than the rest, demonstrating the ongoing activity of a new kind of Web-based threat. These three \pPDF{} \cluster s are large in volume and persistent in time, accounting for \empirical{\seoCampaingPercentage\%} of the total dataset and lasting for the entire duration of our data collection. Also, they exhibit significantly different volumetric and temporal features when compared to email campaigns.
Finally, this paper shows that, while many \pPDF s are distributed as email attachments, the \empirical{three} large \cluster s of \pPDF s are distributed via search engines, exploiting SEO attacks---a new insight placing almost all our files outside the email delivery ecosystem\footnote{While working on this study, Microsoft warned (Tweet: \url{https://twitter.com/MsftSecIntel/status/1403461397283950597}) that the operators of the malware SolarMarker Jupyter are using PDF documents stuffed with SEO keywords to reach victims, further strengthening the importance of our study, indicating a change of distribution strategy.}.

In this paper, we make the following contributions:
\begin{itemize}	
	\item We create and present the first large-scale, pre-labeled dataset of \empirical{\nAnalyzedSamples{}} \pPDF s, featuring \empirical{\nLabels{}} document categories.
	
	\item We identify \empirical{\nMaliciousCampaigns{}} \cluster s out of \empirical{\nLabels{}} whose documents lead to Web attacks.
	
	\item We present the first characterization of \pPDF{} \cluster s, covering different aspects, i.e., volume, duration and activity, visual deceits, and targeted languages.
	
	\item We show that the vast majority of the documents in our dataset is distributed via SEO attacks, i.e., at least \empirical{three} \cluster s, covering \empirical{\seoCampaingPercentage{}\%} of the dataset, with a \empirical{60}-day study of searching \pPDF s on Google and Bing. 
	
	\item We release file hashes, file screenshots, class labels and URLs to the research community.

\end{itemize}

	\section{Background and Methodology} %
\label{sec:research_questions}

Before presenting our study, we define \pPDF{} attacks ($\S$ \ref{sec:prob_stmt}) and outline our methodology ($\S$  \ref{sec:methodology}).

\begin{table}
	\centering
	\footnotesize
	\begin{tabular}{c c c}
		\toprule
		
		\fcolorbox{gray}{white}{\includegraphics[width=0.23\linewidth]{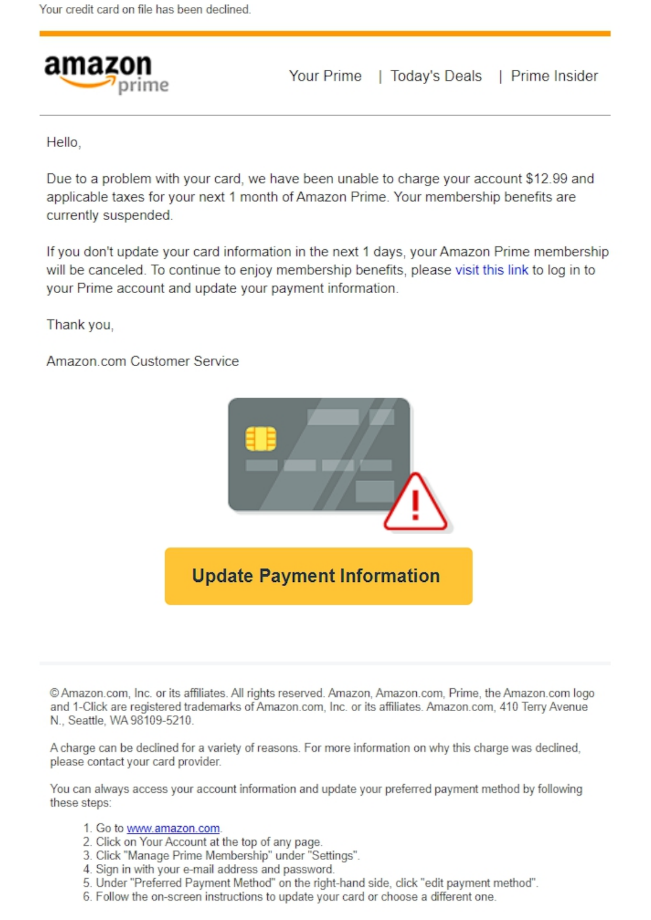}}&

		\fcolorbox{gray}{white}{\includegraphics[width=0.23\linewidth]{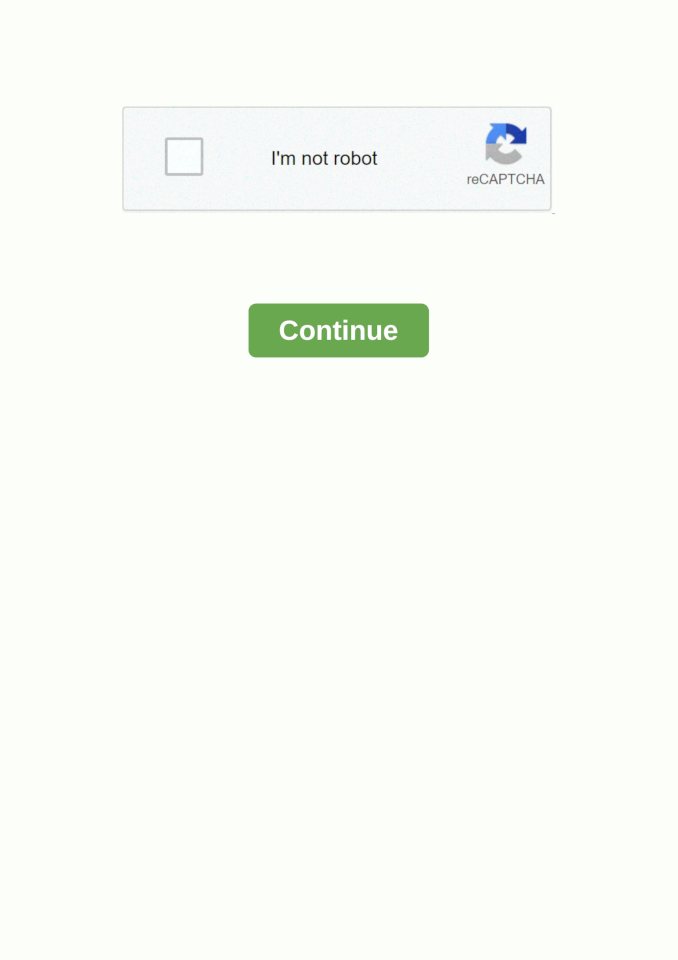}}&

		\fcolorbox{gray}{white}{\includegraphics[width=0.23\linewidth]{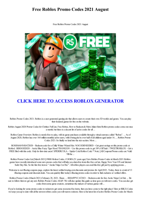}} \\
		
		Amazon message & 
		reCAPTCHA &
		In-game currency\\
		
		\bottomrule
	\end{tabular}
	\captionof{figure}{Examples of \pPDF{} files.}
	\label{fig:examples_background}
\end{table}
\subsection{Background}
\label{sec:prob_stmt}
Previous works discussed PDF files solely as a tool in email phishing attacks, where the deception was in the email body and the exploit occurred via the malicious code embedded in the attached PDF (hereinafter MalPDFs)~\cite{le2017broad, le2014look, thomas2015framing, simoiu2020targeted}. 

Unlike MalPDFs, \pPDF{} files do not embed malware nor do they contain exploits, but they are designed to trick victims into performing an action that can result in landing on malicious web pages that are stealing passwords or user identities, or compromising victims' computers via drive-by downloads~\cite{paloalto_pdftrend, microsoftPhishingAttachments}. \PPDF s rely on a wide variety of visual deceits to lure users into clicking on specific areas of the documents. \Cref{fig:examples_background} shows a few examples of \pPDF s taken from our dataset, using classical phishing patterns, e.g., fake Amazon messages, as well as clickbait messages, e.g., in-game currrency generators.

\subsection{Problem Statement and Methodology}
\label{sec:methodology}
The threat posed by \pPDF s has been object of concern by leading security teams in industry~\cite{microsoftPhishingAttachments, paloalto_pdftrend}.
Despite this anecdotal evidence, the scientific community has largely neglected the threat posed by \pPDF s. %
We follow a strict methodology, performing an array of analyses aimed at providing the first characterization of this phenomenon, based on measurable properties such as volume, activity and duration. We analyze visual baits and structure of \pPDF s looking for signs of diverse exploitation contexts and investigate distribution vectors used by attackers to reach their victims.

Achieving our overarching goal involves addressing both technical challenges and research questions.
First, we tackle the technical challenge of analyzing PDFs at scale.
The characterization of \pPDF s starts with the inspection of the PDFs that our partners receive daily. This daily procedure involves hundreds of documents and is expensive and inefficient, motivating the development of an assistive clustering module. We observe that \pPDF s contain remarkable visual similarities, which we leverage as a clustering feature to drastically reduce the number of PDFs to inspect manually. 
Identifying and enumerating such \cluster s is key to characterize both the general phenomenon and individual \cluster s.

We now turn to our first research question, which requires to \textit{identify and characterize \pPDF s linked to malicious activity.}
We extract all URLs from our PDFs, identifying \textit{bait} URLs---URLs reachable by clicking on visual or textual baits in the first page---that might lead to malicious activity on the Web. We determine maliciousness through a third-party URL analysis service (i.e., VirusTotal) and confirm these results via manual inspection. Next, we focus on those \cluster s whose \pPDF s evidently lead to attacks on the Web and proceed with their characterization.
Our analysis focuses first on measurable properties, such as \cluster{} size, duration, activity and temporal dynamics (similarly to prior works, e.g.,~\cite{simoiu2020targeted,van2019cognitive, irani2008evolutionary}) as well as their reach, by measuring the number and distribution of languages across and within \cluster s.  %
Additionally, we discuss the visual baits of \pPDF s, searching for indications of attackers' reliance on different exploitation contexts other than the email distribution ecosystem (e.g., Web).

Then, we \textit{investigate two possible distribution vectors.}
Understanding the origin of \pPDF s is a key component in characterizing this phenomenon.
Previous works only discussed PDFs as part of email phishing campaigns. We quantify how many \cluster s are distributed as email attachments by matching files on a corporate spam trap and by leveraging VirusTotal metadata.
Beyond that, empirical observations on the structure of \pPDF s suggest another distribution mean: the documents of the three largest \cluster s share the common traits of Search Engine Optimization (SEO) attacks, i.e., keyword stuffing~\cite{webspam}, cross-linking resources~\cite{linkfarm}, and use of benign websites for linked resources~\cite{deseo}. We hypothesize that attackers poison search engine results to increase the visibility of these files to reach their victims. 
We verify this hypothesis by inspecting search results of popular search engines, such as Google and Bing~\cite{SearchEn9:online}, for 30 days. %

	\section{Dataset and \cluster s} %

Our analysis relies on a dataset of \nAnalyzedSamples{} PDF documents with unique SHA256 signature, collected from Dec. 16th, 2020 to Jun. 23rd, 2021.
In this section, we describe the sources of data and data collection procedures ($\S$ \ref{sec:dataset}). Then, we report the procedure we followed to extract clusters of visually similar documents ($\S$  \ref{sec:campagin_identification}).

\subsection{Dataset}
\label{sec:dataset}

\paragraph*{Data Sources.}

The sources of our dataset are two industrial partners, i.e., \PartnerA{} and \PartnerB{}\footnote{The names of the two partner companies are anonymized at submission time, but we can provide the names after consulting with the PC chairs. \PartnerA{} is a global corporation in the field of networks, telecommunications, and security, with a number of employees in the order of tens of thousands. \PartnerB{} is a SME in the field of packet inspection, network security, and threat intelligence.}, who  provided us with daily feeds of PDF files. \PartnerA{} started sending us data on Dec. 16th, 2020.
To increase the diversity and coverage of the dataset, we introduced a second industrial partner, \PartnerB{}, starting from Mar. 3rd, 2021. We were concerned that \PartnerA{}'s sampling policy regarding the least number of AV flags (see $\S$ \textit{Data Collection} below) might have introduced a bias towards documents with a higher number of AV flags. We sought to counter-balance this effect by including documents with lower AV scores, as a minimum threshold was not imposed by \PartnerB{}.
\Cref{fig:samples-over-time} shows daily uploads aggregated per week until the end of this study, Jun. 23rd, 2021, highlighting the contribution of each partner; the respective areas are stacked to highlight the total weekly amount. The contribution of \PartnerA{} and \PartnerB{} to the dataset is of \empirical{\contribCisco}\% and \empirical{\contribInquest}\%, respectively, with a negligible fraction of shared samples over the total, i.e., \empirical{\sharedCisInq}\%.

\paragraph*{Data Collection.}
\PartnerA{} retrieves data from VirusTotal~\cite{VirusTot72:online} (VT), fetching PDF files uploaded on the previous day and flagged as malicious by at least nine antivirus (AV) engines by using search modifiers, a VT feature to filter files on properties such as file type, size, and the number of engines flagging the file as malicious. \PartnerB{} receives feeds of malicious documents from multiple sources, one of which is VT, and shares with us those samples which are also confirmed from a second source. \PartnerB{} retrieves samples from VT using selectors specified via YARA rules~\cite{yara}, a rule-based approach designed for the description of malicious files. \PartnerB's rules search for unseen PDFs tagged as phishing, flagged as malicious by at least one AV engine, with encrypted PDF objects, or tagged with embedded JavaScript. The list of the rules used by \PartnerB{} is publicly available~\cite{yarainquest}. %
We receive samples from \PartnerB{} on the day they are uploaded on VirusTotal.

\paragraph*{Data Preprocessing.} 
At first, we rule out the possibility that our dataset contains PDFs with exploits or malicious JavaScript. We look for PDFs tagged by VT with \texttt{js-embedded}, \texttt{file-embedded}, \texttt{exploit}, \texttt{cve-xxxx}, and \texttt{launch-action}, which indicate the presence of exploit code or malware, and find that MalPDFs are a negligible fraction of our dataset (\empirical{\numprint{0.24}\%} or \empirical{440} files).

\newcommand{\sahar}[1]{\textcolor[rgb]{0,0,0}{#1}}
\subsection{PDF Clustering}
\label{sec:campagin_identification}

The first challenge we address is grouping PDF documents using an appropriate similarity metric. 
As exhaustively inspecting all documents manually is not scalable, our goal is to implement a procedure for grouping documents whose content is visually similar, with the aim of using this by-product to speed up human inspection of the daily PDF feed.
A common clustering approach for phishing messages relies on Natural Language Processing (NLP), where the similarity metric is calculated using the text in the message (e.g.,~\cite{simoiu2020targeted,van2019cognitive, irani2008evolutionary}). However, PDF documents in our dataset do not exclusively rely on text to convey the fraudulent message, e.g., the fake reCAPTCHA documents, making it challenging for NLP-based clustering to produce meaningful clusters. Another approach to determine document similarity is by using raw document screenshots and supervised learning (e.g.,~\cite{abdelnabi2020visualphishnet}). Unfortunately, supervised learning techniques rely on a pre-existing labeled training set, which is unavailable in our case, making supervised learning unsuitable for our goal. We thus resort to unsupervised learning techniques to assist the identification of \cluster s of visually-similar PDF files.

\paragraph*{Clustering.}
\label{sec:assisted_clustering}

Previous work shows that replacing raw images with Convolutional Neural Networks (CNNs) features can lead to better clustering performance~\cite{guerin2017cnn, guerin2018improving}. Thus, we utilize the DeepCluster framework~\cite{caron2018deep}, a recent work in unsupervised representation learning, that jointly trains a CNN with $k$-means clustering. In each epoch, the training alternates between training the CNN and clustering and computing the pseudo-cluster-labels. We adopt the same DeepCluster setup (AlexNet architecture~\cite{krizhevsky2012imagenet}) with mainly two changes: (i) We keep color information, as it can be a distinguishing factor; (ii) We decrease the number of clusters from \numprint{10000} to 900, as we have a smaller dataset with a lower expected number of clusters.

We generate a raw screenshot of the first page of a PDF using \texttt{pdftoppm}~\cite{poppler} with 150 dots per inch (DPI) and obtain \nAnalyzedSamples{} screenshots.
As a pre-processing step, we remove images  with the same p-hash value (obtained from documents with different SHA256), lowering the number of samples to \nSamplesAfterPhash{}.
Once we trained and ran DeepCluster on the screenshots with unique p-hash values, we validate the 900 clusters by randomly selecting 10 documents per cluster (\numprint{9000} samples in total) and determining the screenshot similarity considering text and image positions. As an output of this step, we identify \empirical{635} homogeneous clusters covering \empirical{\nSamplesHomogeneousAfterDC} (\empirical{90\%}) of the input samples. This clustering step split large \cluster s into many smaller, fine-grained ones, therefore we merge homogeneous clusters containing similar documents. At the end of this step, we obtain \empirical{15} distinct \cluster s of documents. %

To cluster similar documents in the remaining \empirical{\nDBSCANSamples} (\empirical{10\%}) samples, we run DBSCAN~\cite{ester1996density}, using the learnt embeddings as distance metric (as in, e.g.,~\cite{caron2018deep}): we obtain  \empirical{\nDBSCANClusters} clusters and \empirical{\nDBSCANnoise} noise points.
We subsequently confirm that \empirical{\nDBSCANHomCl} clusters (610 samples) of the \empirical{\nDBSCANClusters{}} are homogeneous and identify \empirical{29} new \cluster s obtained by merging similar homogeneous clusters. %
As a refinement step, we manually cluster the remaining \empirical{\numprint{1504}} documents, discovering another 36 \cluster s, and group \empirical{389} spurious documents in the \textit{Outliers} \cluster{}. 
\Cref{tab:camp_ident_size} (Appendix) reports the amount of documents involved at each clustering step.
Finally, we assign each cluster an arbitrary name of our choice, with the only purpose of helping the authors remember the outlook of each of them, and redistribute the \empirical{\nPhashRemoved{}} samples that we filtered out by means of perceptual hash, assigning them to the cluster of their matching sample.
The final number of PDF \cluster s observed in the dataset is \empirical{\nLabels{}}, including \textit{Outliers}. The interested reader can find more details on the clustering procedure and validation in \Cref{sec:clustering_details}.

\section{Establishing Maliciousness}
\label{sec:attacks}

PDF documents, including \pPDF s, may contain URLs in any page. More importantly, \pPDF s exhibit the specific feature of embedding a URL leading to a Web attack in the first page.
We use the presence of such \textit{malicious} URLs as a discriminating factor to identify \pPDF s.
In this section, we first present the extraction methodology for the URLs embedded in all \nAnalyzedSamples{} documents.
Then, we identify PDFs linked to an ongoing malicious activity on the Web. Finally, we detail the observed attacks and motivate the soundness of our findings.

%
%
%

\subsection{URL Extraction}
\label{sec:url_extr_analysis}
Although trivial at a first glance, URL extraction from \pPDF s poses a few challenges.
First, PDF files can contain encoded (e.g., base 64), compressed (e.g., deflate), or encrypted objects and streams, removing the string markers characterizing URLs, such as \texttt{http://}.
Next, automated PDF generation from attackers may lead to corrupted or invalid permutations of the PDF structure where, e.g., URL-bearing PDF objects are disconnected from the PDF graph and thus not clickable, or they have a null clickable area.
Below, we detail the URL extraction procedure, which ensures the extraction of clickable, well-formed first-page URLs (\textit{bait} URLs) at scale.

We produce a normalized representation of each PDF file by removing any encoding or compression. Decrypting streams and objects was not possible because we did not have the encryption key. Then, we extract a graph-like representation of the normalized PDF with \texttt{peepdf}~\cite{peepdf}, a popular tool for analyzing malicious PDFs.
We traverse the graph-like structure starting from the root element (the \texttt{Catalog} node) using a breadth-first algorithm to avoid loops, searching for those nodes containing links. Using regular expressions to extract URL-looking string text may increase the number of false positives. Accordingly, we leverage the semantic of the graph-like structure, searching for the PDF elements used to implement document areas that result in visiting a URL upon a mouse click. Such an area is implemented as a node containing a \texttt{URI} node having ancestors with the attribute \texttt{Subtype Link}, the attribute \texttt{Rect}, and either the \texttt{Type Annot} or \texttt{Type A} attribute. 
Further, we remove ill-formed URLs (e.g., the top-level domain is invalid, the URL network location is \texttt{127.0.0.1} or the URL scheme is not HTTP or HTTPS) and URLs pointing to static resources such as images or JSON files, which do not present a threat to users. 

\begin{figure}
	\centering
	\includegraphics[width=0.99\linewidth]{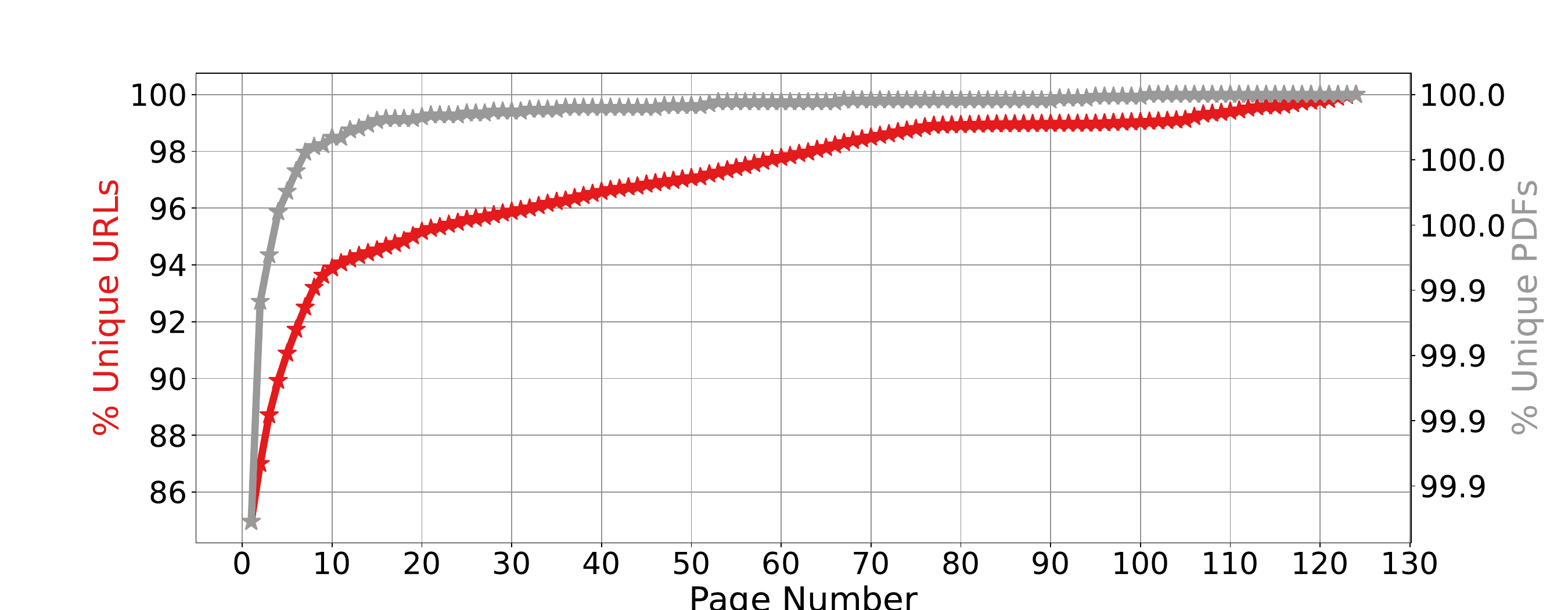} 
	\caption{Distribution of \textit{bait URLs} per PDF page (red) and number of unique PDFs embedding them (grey). The graph shows the .95 quantile of PDF pages (max: 524) for visibility reasons.}
	\label{fig:baitURL_distrib}
\end{figure}

We verify that PDFs in our dataset are more likely to include links in the first page rather than in following pages by plotting the distribution of unique \textit{bait URLs} per page, shown in red in \Cref{fig:baitURL_distrib}.
We observe that \empirical{86\%} of all the extracted URLs are first-page URLs, covering \empirical{99\%} of the PDFs.
This distribution confirms our intuition that first-page URLs are relevant features of our PDFs and that they are worth analyzing. First-page URLs, being displayed first to victim users, are more likely to lead to an attack.
We thus discard URLs in pages after the first, obtain \empirical{\distinctPageOne{}} unique URLs, and focus the next steps of our analysis on first-page bait links.


\subsection{URL Analysis}
\label{sec:url_valid}

After the extraction step, we determine which URL points to a malicious webpage.
A common technique to determine the maliciousness of URLs is using URL blocklists, such as Google Safe Browsing (GSB)~\cite{SafeBrow41:online}. Blocklists like GSB intend to offer a live protection mechanism for browsers to warn users visiting a malicious website at the time of the visit. As a result, URLs that are no longer malicious or no longer exist are evicted from the blocklist, reducing our ability to determine maliciousness after a short period of time. 
We empirically observed that in some cases the time interval between the start of the malicious activity of a webpage and our reception of the PDF via VirusTotal is non-negligible, especially when considering web attacks such as phishing, whose malicious activities last on average 21 hours~\cite{oest2020sunrise}.
Such malicious bait links might already be offline or evicted from the blocklist by the time we look them up.
A better option for our case study is using URL analysis services with historical data, e.g., VirusTotal or \texttt{urlscan}\cite{urlscan:online}.
Thanks to \PartnerA{}'s availability of 20K URL analysis requests on VT, we randomly sampled an equal number of URLs from each \cluster{}, until either the entire \cluster{} was covered or the cap was reached.
To ensure validity of our approach, we inspected its coverage by \cluster{}. 
Our sampling offers a high coverage, of \empirical{100\%} for all \cluster s except for 14, where we covered from \empirical{1.28\%} (or \empirical{\numprint{1000}} files) of the \textit{reCAPTCHA} \cluster{} up to \empirical{99.69\%} (or \empirical{765} files) of the \textit{NSFW `Find'} \cluster{}. \Cref{tab:vt_coverage} (Appendix) shows the coverage per \cluster{}. 

We also perform a manual inspection of \empirical{722} randomly-sampled first-page well-formed clickable URLs (\textit{bait links}) to determine maliciousness. 
We label a URL as malicious if we observe any of the following behaviours: prompting file download, user interaction (click), asking for permissions, modifying the browser settings, leading to a phishing page, a Google SafeBrowsing warning, or to other types of unwanted content. Otherwise, we label the URL as benign. 

\begin{table}
	\centering
	\footnotesize
	{
		\begin{tabular}{l | c r r r  | r r r }
			\toprule
			\myalign{c}{\rot{90}{\shortstack{\Cluster{}\\Identifier}}}               & \myalign{c}{\rot{90}{\shortstack{Attack\\Type}}}               & \myalign{c}{\rot{90}{Volume}}       & \myalign{c}{\rot{90}{\shortstack{\# Unique\\Phash}}} & \myalign{c}{\rot{90}{Avg/day}} &  \myalign{c}{\rot{90}{First seen}} & \myalign{c}{\rot{90}{Last seen}}  &                \myalign{c}{\rot{90}{\% Active}} \\
			\midrule			
			reCAPTCHA                           & \Circle     & \numprint{78854}  &              157&    436 &    16.12.20 & 23.06.21 &                     95.8\% \\
			ROBLOX Text                         & \Circle     &  \numprint{59348} & \numprint{16399}&    667 &    06.03.21 & 23.06.21 &                     81.7\% \\
			\textit{ROBLOX Picture}             & \Circle     &  \numprint{18065} &             192 &    278 &    05.03.21 & 23.06.21 &                     59.1\% \\
			NSFW `Play'                         & \varhexstar & \numprint{9797} &               274 &     55 &    17.12.20 & 23.06.21 &                     94.7\% \\
			\textit{reCAPTCHA Drive}            & \Circle     &   \numprint{1693} &              15 &     18 &    12.02.21 & 23.06.21 &                     73.3\% \\
			
			\textit{Download Torrent}           & \Circle     &   \numprint{1121} &             112 &     18 &    15.02.21 & 23.06.21 &                     48.4\% \\
			Ebooks                              & \Circle     &    795 &                        458 &         7 &    17.12.20 & 22.06.21 &                     61.5\% \\
			NSFW `Find'                         & \DOWNarrow  &    322 &                         45 &         4 &    20.01.21 & 20.06.21 &                     58.3\% \\
			CLICK-HERE                          & \Circle     &    286 &                         58 &         3 &    09.03.21 & 21.06.21 &                     81.7\% \\
			PDF Blurred                         & \CIRCLE     &    228 &                         27 &         3 &    11.01.21 & 23.06.21 &                     44.2\% \\
			Coin Generator                      & \Circle     &    167 &                        115 &         3 &    23.12.20 & 23.06.21 &                     28.0\% \\
			Russian Forum                       & \Circle     &    167 &                         12 &         3 &    23.12.20 & 21.06.21 &                     29.4\% \\
			AS PDF / File \#1                   & \CIRCLE     &    134 &                         17 &         2 &    24.12.20 & 22.06.21 &                     40.0\% \\
			Elon Musk BTC                       & \Circle     &     82 &                         17 &         4 &    06.02.21 & 22.06.21 &                     14.7\% \\
			Try Your Luck                       & \UParrow    &     79 &                         25 &         7 &    29.12.20 & 17.06.21 &                      6.5\% \\
			Play Video                          & \Circle     &     70 &                         56 &         2 &    05.03.21 & 22.06.21 &                     38.5\% \\
			\textit{Access Online Gen.}         & \Circle     &     55 &                         6 &         4 &    20.12.20 & 04.05.21 &                      9.6\% \\
			NSFW `Click'                        & \UParrow    &     44 &                         15 &         3 &    12.02.21 & 02.06.21 &                     11.8\% \\
			Lottery 25th Ann.                   & \UParrow     &     43 &                         23 &         2 &    19.01.21 & 28.05.21 &                     20.2\% \\
			AS PDF / File \#4                   & \CIRCLE     &     41 &                         12 &         1 &    23.12.20 & 04.06.21 &                     18.4\% \\
			Apple receipts                      & \CIRCLE     &     30 &                         21 &         1 &    20.12.20 & 11.06.21 &                     15.6\% \\
			Download Btn                        & \Circle     &     19 &                         19 &         1 &    19.12.20 & 26.05.21 &                     11.4\% \\
			Fake SE                             & \Circle     &     18 &                         17 &         1 &    01.02.21 & 05.05.21 &                     19.4\% \\
			Amazon scam                         & \DOWNarrow  &     14 &                         11 &      1 &    20.01.21 & 11.06.21 &                      8.5\% \\
			NSFW `Dating'                       & \DOWNarrow  &     14 &                         13 &      5 &    17.04.21 & 07.06.21 &                      5.9\% \\
			Download PDF                        & $\Diamond$  &     13 &                         13 &      1 &    14.02.21 & 17.06.21 &                     10.6\% \\
			AS PDF / File \#11                  & \Square + \DOWNarrow & 11 &                     6 &         1 &    03.02.21 & 08.06.21 &                      8.8\% \\
			AS PDF / File \#3                   & \Square     &     11 &                          7 &      1 &    11.03.21 & 25.05.21 &                     10.7\% \\
			\textit{Sigue Leyendo}              & \XBox       &     10 &                          7 &      1 &    27.02.21 & 03.06.21 &                     10.4\% \\
			Web Notification                    & \Square     &      8 &                          2 &      1 &    10.03.21 & 04.05.21 &                     12.7\% \\
			Link farm                           & \UParrow    &      7 &                          6 &      2 &    17.01.21 & 04.04.21 &                      5.2\% \\
			\textit{AS PDF / File \#10}         & \Square     &      6 &                          3 &      1 &    26.12.20 & 07.06.21 &                      3.7\% \\
			\textit{AS PDF / File \#8}          & \Square     &      6 &                          4 &      1 &    25.03.21 & 14.04.21 &                     25.0\% \\
			AS PDF / File \#6                   & \CIRCLE     &      5 &                          2 &      1 &    18.03.21 & 03.06.21 &                      6.5\% \\
			Netflix scam                        & \UParrow    &      5 &                          2 &      3 &    21.12.20 & 23.12.20 &                    100.0\% \\
			Get Your Files                      & \CIRCLE     &      4 &                          2 &      1 &    10.03.21 & 17.03.21 &                     42.9\% \\
			QR code                             & \CIRCLE     &      3 &                          3 &      2 &    21.01.21 & 22.03.21 &                      3.3\% \\
			\textit{Click Here TShirt}          & \Circle     &      3 &                          3 &      1 &    26.03.21 & 17.04.21 &                     13.6\% \\
			\textit{Download File}              & \Circle     &      3 &                          3 &      2 &    03.09.21 & 21.05.21 &                      2.7\% \\
			AS PDF / File \#7                   & \CIRCLE     &      3 &                          3 &      1 &    16.04.21 & 18.05.21 &                      9.4\% \\
			AS PDF / File \#13                  & \DOWNarrow  &      2 &                          2 &      1 &    10.02.21 & 07.06.21 &                      1.7\% \\
			Adobe Click                         & \CIRCLE     &      2 &                          2 &      1 &    26.01.21 & 09.06.21 &                      1.5\% \\
			SharePoint                          & \CIRCLE     &      2 &                          2 &      1 &    04.05.21 & 02.06.21 &                      6.9\% \\
			Shared Excel                        & \UParrow    &      2 &                          2 &      1 &    06.01.21 & 12.02.21 &                      5.4\% \\
			\bottomrule 
	\end{tabular}}
	\caption{The \empirical{\nMaliciousCampaigns{}} \cluster s associated with malicious activity. \Cluster s in italics were validated by manual inspection only. Dates are in \texttt{dd.mm.yy} format.}
	\label{tab:all_campaigns}
\end{table}

\subsection{Observed Malicious Activity} 
\label{sec:mal_results}
\PartnerA{} fetched a total of \empirical{\nUrlsSubmVt} distinct URL reports, where \empirical{89\%} of the URLs were unknown to VirusTotal, \empirical{7\%} were flagged as benign, and \empirical{4\%} (\empirical{868}) were flagged as malicious. The reasons behind this low number of URLs known to VT are unclear to us, and studying the AV inner workings goes beyond the scope of our research questions. We empirically observed that VT may have no knowledge of links embedded in PDFs even when one or more of its partner AVs flags the binary file as malicious. We discuss this observation in $\S$ \ref{sec:discussion_VT_coverage}.
The \empirical{868}  malicious URLs flagged by VirusTotal belong to \empirical{52} \cluster s. 
Our manual analysis validated both URLs that were labelled as malicious (\empirical{32\%} of the manually-analyzed URLs) and URLs that were flagged as benign or never scanned (\empirical{61\%}), and confirms \empirical{44} of the malicious \cluster s reported by VT.
Conversely, we observed that URLs belonging to \empirical{eight} \cluster s were not malicious, containing documents about phishing training, generic text documents or ebooks, invoices, articles about security, reports by a security firm, flyers about events, or screenshots of a tool by Netcraft. Further details are reported in \Cref{sec:fp_benign_categories}.
The manual analysis also flagged URLs, not flagged by VT and belonging to \empirical{nine} \cluster s, as malicious, and identified benign URLs belonging to \empirical{five} \cluster s.

Overall, the URL analysis returned eight different outcomes, reported in \Cref{tab:all_campaigns} and detailed in the following.
\textit{Malicious advertisement and Data harvesting (\empirical{16} \cluster s, symbol: \Circle)}: in this attack, the user is redirected to a personalized advertisement page or is prompted to provide personal data to receive a reward (similarly to, e.g.,~\cite{kharraz2018surveylance}). 
\textit{Google SafeBrowsing warnings (\empirical{10} \cluster s, symbol: \CIRCLE)}: GSB warned against either phishing or harmful content.
\textit{Malware (\empirical{five} \cluster s, symbol: \Square)}: the web page prompts to download a file (e.g., Office documents) or suggests to install additional software. We observed \empirical{one} \cluster{} delivering multiple attacks and classified it accordingly.
\textit{Phishing (\empirical{four} \cluster s, symbol: \DOWNarrow)}: these pages delivered classic phishing attacks.
\textit{VirusTotal (\empirical{six} \cluster s, symbol: \UParrow)}: the evidence of malicious activity was provided by VirusTotal results.
\textit{Various attacks (\empirical{three} \cluster s)}, which include: \textit{Drugs promotion} (symbol: \XBox), where one \cluster{} led to a blog promoting diet pills; \textit{Fake search engine} (symbol: $\Diamond$), describing one \cluster{} leading to a page pretending to be a search engine; \textit{Adult content (symbol: \varhexstar)}, describing one \cluster{} leading to an adult website.

\subsection{Summary of Findings}
The goal of this section was to analyze representative URL samples for all the \cluster s obtained in $\S$  \ref{sec:campagin_identification}, investigating whether these URLs lead to Web attacks.
In \empirical{\nMaliciousCampaigns{}} \cluster s all analyzed active URLs led to an attack webpage, where the attack types are consistent. This pattern of homogeneity in attack types among the \cluster s suggests that they may be linked to malicious activity.
Conversely, URLs from \empirical{nine} other clusters showed signs of malicious activity as well as of benign activity (at least one malicious and one benign URL). We excluded them from the rest of the analyses, as we conservatively select \cluster s linked to malicious activity only.

	\section{\Cluster s Characterization} %
\label{sec:campaignAnalysis}

We now characterize each of the \empirical{\nMaliciousCampaigns{}} \cluster s identified in $\S$ \ref{sec:mal_results}. First, we look at volumetric and temporal properties of each \cluster{} ($\S$ \ref{sec:vol_temp}). Second, we analyze the visual deceits of each \cluster{} ($\S$ \ref{sec:visual_deceits}), providing a categorization of the type of fraudulent activities and their visual elements. Then, we explore the effectiveness of the VirusTotal maliciousness score ($\S$ \ref{sec:virustotal_score}). Finally, we study the geographical reach of each \cluster{} by observing the languages used in their text ($\S$ \ref{sec:targetAnalysis}). 

\begin{figure*}[h]
	\centering
	\begin{subfigure}{0.3\linewidth}
		\centering
		\includegraphics[width=0.95\linewidth]{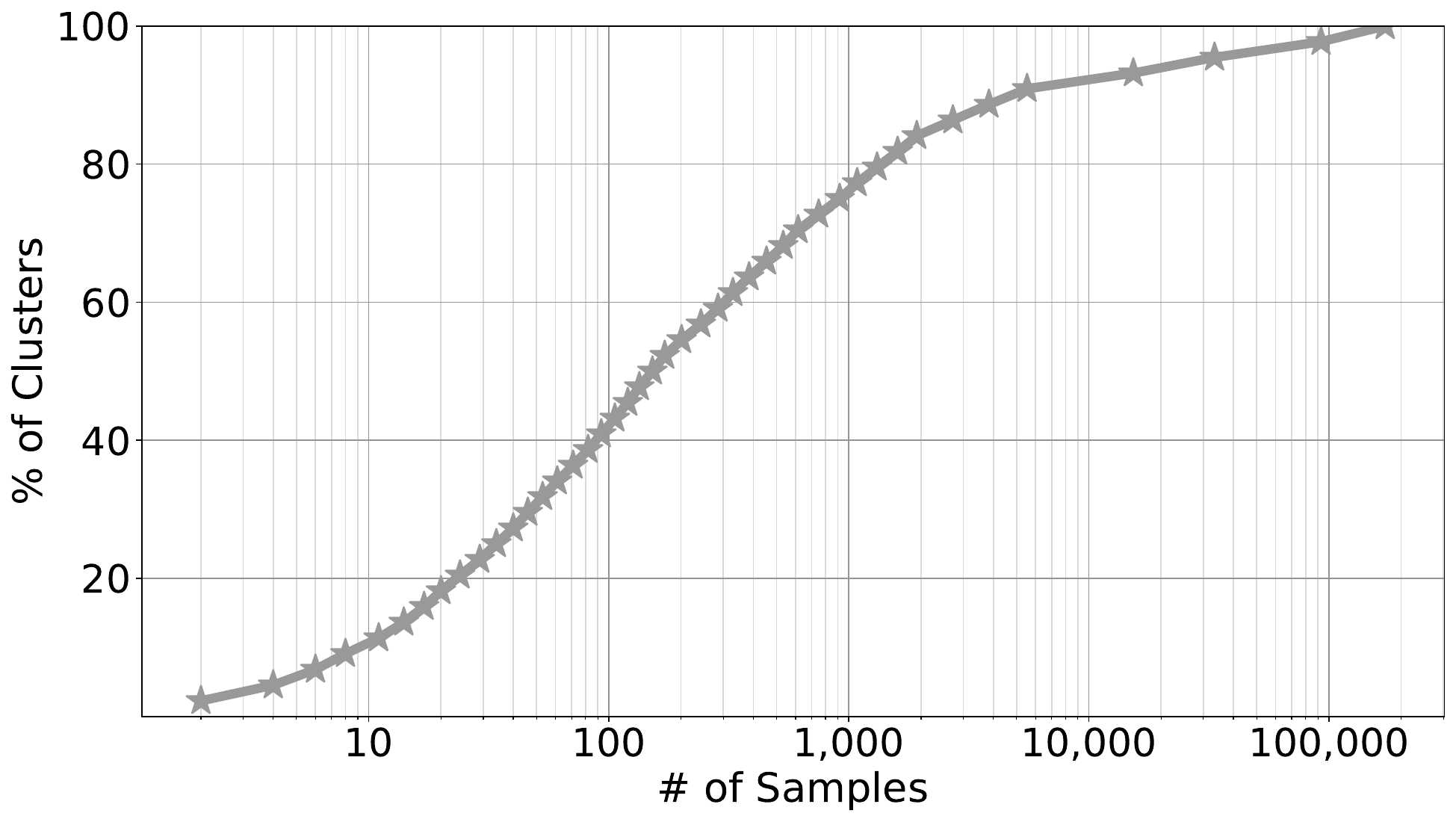}
		\caption{}
		\label{fig:cdf_volume}
	\end{subfigure}
	~~
	\begin{subfigure}{0.3\linewidth}
		\centering
		\includegraphics[width=0.95\linewidth]{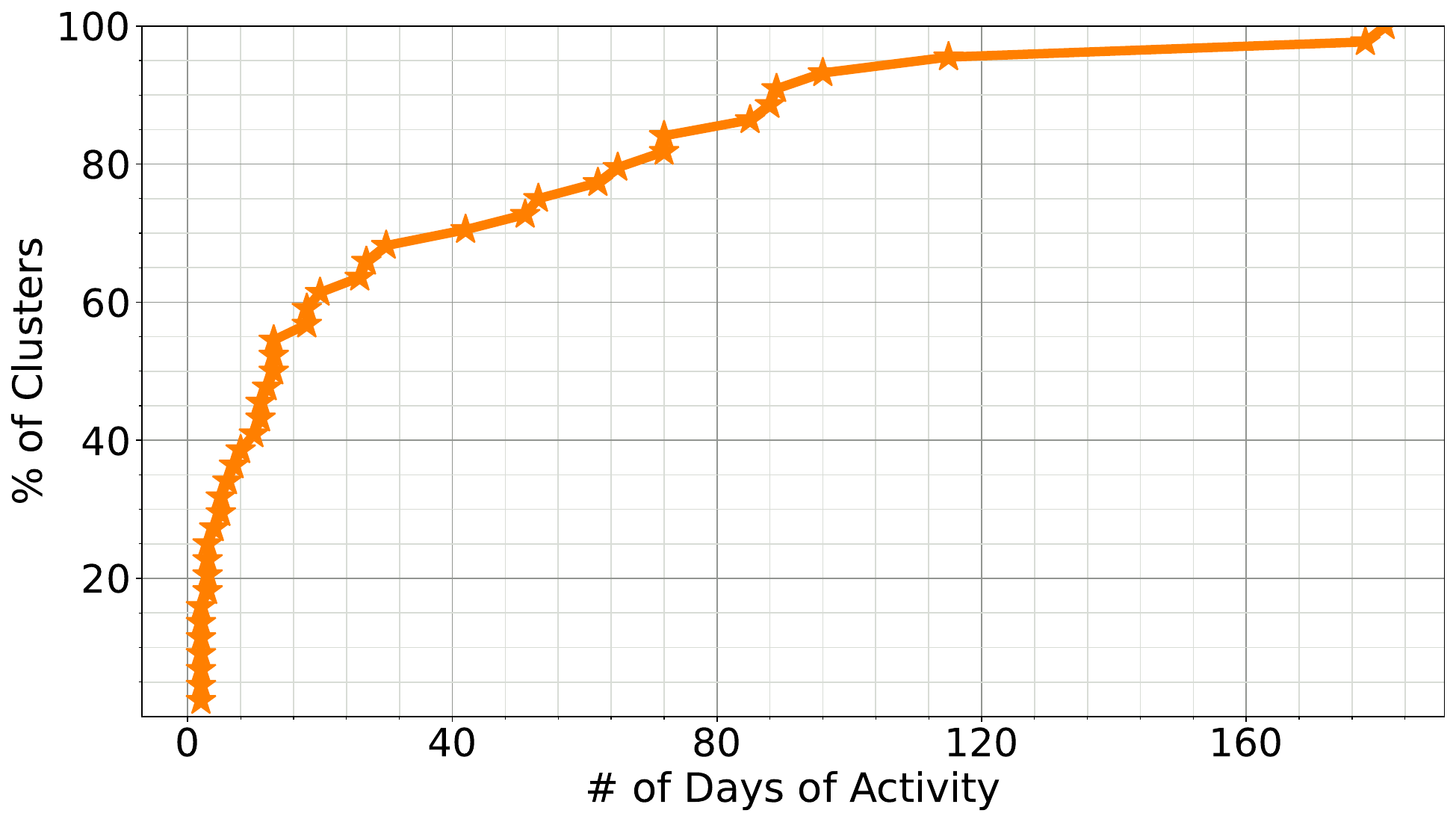}
		\caption{}
		\label{fig:cdf_activity}
	\end{subfigure}
	~~
	\begin{subfigure}{0.3\linewidth}
		\centering
		\includegraphics[width=0.95\linewidth]{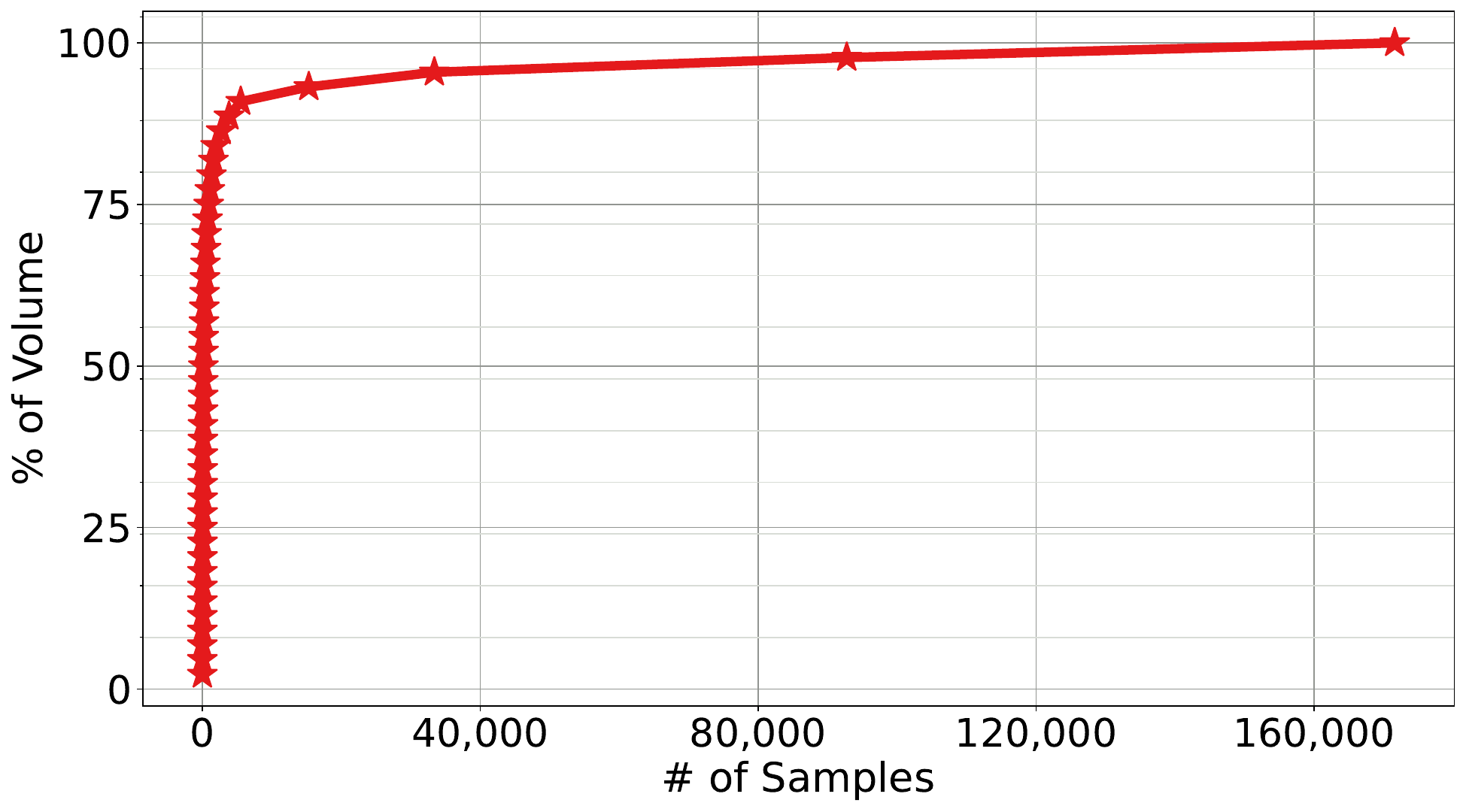}
		\caption{}
		\label{fig:cdf_activity_vol}
	\end{subfigure}
	\begin{subfigure}{1\linewidth}
		\centering
		\includegraphics[width=\textwidth]{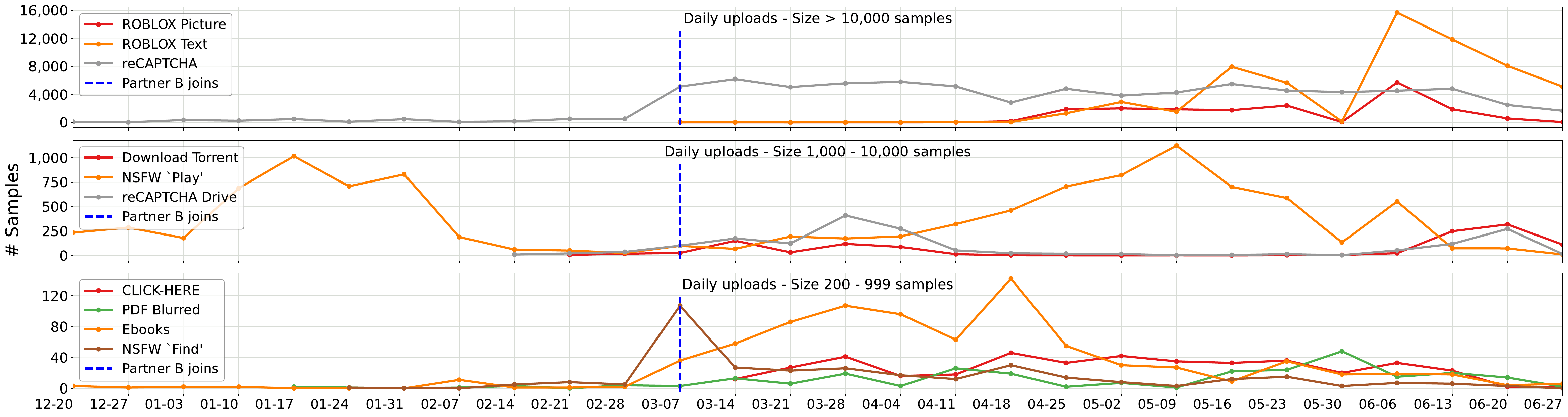}
		\caption{\Cluster s volumes grouped by \cluster{} over time.}
		\label{fig:campaign_over_time}
	\end{subfigure}%
	\caption{Cumulative Distribution Function of:  (a) The volume of \pPDF{} documents over number of \cluster s. (b) The \cluster{} activity in days over number of \cluster s. (c) The contribution of \cluster{} volumes over the total dataset.}
\end{figure*}

\subsection{Volumetric and Temporal Dynamics}
\label{sec:vol_temp}

\paragraph{Volume.}
\label{sec:size_and_volume}

\PPDF{} files are not evenly distributed over the \empirical{\nMaliciousCampaigns{}} malicious \cluster s. \Cluster{} sizes are skewed, with the top \empirical{5}\% of malicious \cluster s (i.e., \empirical{three} \cluster s) corresponding to about \empirical{\seoCampaingPercentage{}}\% of the dataset, while \empirical{78\%} of the \cluster s contain fewer than \numprint{1000} documents and \empirical{42\%} contain fewer than 100 (see \Cref{fig:cdf_volume} and \ref{fig:cdf_activity_vol}).

\paragraph{Duration and Activity.}
\label{sec:duration_and_activity}

The temporal dynamics of the \cluster s are diverse. For example, \cluster s like \textit{reCAPTCHA} tend to be constant, without notable peaks. We speculate that the absence of patterns and peaks may indicate that their discovery and upload on VirusTotal may be automated. In contrast, other \cluster s, e.g., the two \textit{ROBLOX} \cluster s, all \cluster s with sizes between \numprint{1000} and \numprint{10000} samples, and \textit{NSFW `Find'}, have a less regular evolution, indicating periods of low and high activity. \Cref{fig:campaign_over_time} shows the temporal dynamic of the \cluster s by number of daily uploads, grouped by the total size of the \cluster{} (200 - 999 samples, \numprint{1000} - \numprint{10000} samples, and more than \numprint{10000} samples).

We observe that most clusters are active for a period between one and two months, where specifically \empirical{28\%} of them are active for up to five days and \empirical{77\%} of them are active for at most 60 days (see \Cref{fig:cdf_activity}). While few \cluster s operate for 60 days or more (\empirical{11} \cluster s), their total size covers \empirical{99\%} of the entire dataset, with \empirical{three} \cluster s lasting more than \empirical{100} days (i.e., \textit{reCAPTCHA}, \textit{NSFW `Play'} and \textit{Ebooks}).
These activity periods are considerably long, especially in comparison with email-based phishing campaigns, which last one day on average~\cite{simoiu2020targeted}.
\Cref{tab:all_campaigns} shows size, prevalence, duration and temporal location for all the \empirical{\nMaliciousCampaigns{}} malicious \cluster s.

\subsection{Visual Deceits} 
\label{sec:visual_deceits}

Attackers use visual deceits to lure victims into clicking~\cite{blythe2011f, dhamija2006phishing}. We enumerated the types of visual baits and clickbait messages conveyed by the document text and identified two types of deceits. If a document includes logos, images or phrases reproducing existing entities (e.g., a company), processes (e.g., sharing of a document) or situations (e.g., receiving a money transfer), we categorize it as \textit{Impersonation}. Otherwise, when a document entices the victim into clicking in order to obtain paid goods, illegal goods, or other unwanted content (e.g., adult content), we categorize it as \textit{Promotion}. 
Also, we consider whether visual elements in \pPDF s may be similar to those found in different contexts. In particular, we look for PDFs resembling invoices, cloud or email notifications, and documents with UI elements used in web pages.
The \cluster s distribute evenly between the two types of deceit.

\paragraph*{Promotion.}  Promotion \cluster s can be further divided into four sub-\cluster s: in-game currencies or pirated content (\empirical{15} \cluster s), material goods, e.g. electronic devices or money (\empirical{two} \cluster s), adult content (\empirical{four} \cluster s), and drugs (\empirical{one} \cluster{}).
With \empirical{two} large-size \cluster s, this deceit category covers \empirical{45}\% of the dataset.

The layout of these documents is usually not elaborate: \empirical{64}\% of them have a bare structure including an image for the advertised product, a catchphrase or bait (e.g., ``Click here for free BTC'') and a button, \empirical{18}\% are very text-heavy, employing techniques such as keyword stuffing and randomization, and \empirical{five} \cluster s show with varying levels of detail renown visual elements such as video players, hubs for content sharing, or threaded discussions.

\paragraph*{Impersonation.} The \cluster s in this category disguise their content as legit, mimicking existing commercial services, communications or people by means of typographic and visual elements, and ask to review the status of a process, access a shared document or prove their identity. 
In \empirical{17} cases documents reproduce parts of communications (e.g., emails from colleagues, friends or firms) or behaviors of viewer programs, prompting for valid credentials to access a protected file.
In the remaining cases, the documents mimic established and widely recognized Web UI components or processes, like search engine results or CAPTCHA challenges by including key textual and graphical elements. For example, they display search results on the initial PDF page just like in a web browser, feature a reCAPTCHAv2 challenge image at the center of the first page, or show a browser popup requesting permissions. We note that attackers overlay large clickable areas around them.
These UI elements are familiar and linked to authentic services, which operate by briefly halting user interaction with the page until they are removed with a click. %
\Cluster s displaying such visual baits likely exploit the notion that such an interruption is inconspicuous, as it aligns with typical behavior, and can be dismissed through a click.
This characteristic of \pPDF s strikes a difference from conventional attack scenarios focused on attachments, opening up alternative possibilities such as employing the PDF as an intermediary step within a redirection chain.

\subsection{VirusTotal Score for Maliciousness}
\label{sec:virustotal_score}

\begin{figure}
	\centering
	\includegraphics[width=0.99\linewidth]{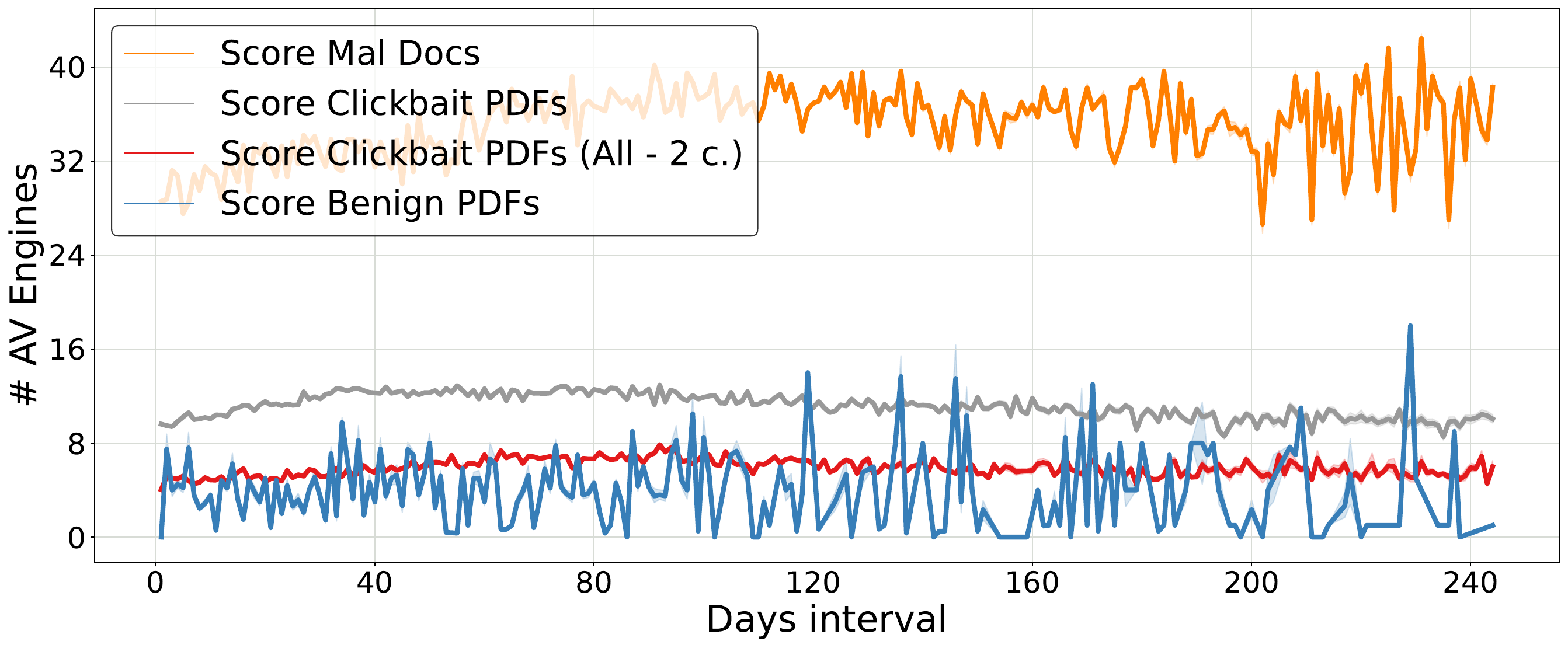} %
	\caption{VirusTotal score comparison between MalDocs and \pPDF s. Data collection until Aug, 18th.}
	\label{fig:vt_score}
\end{figure}

Prior studies on malware programs have relied on the VirusTotal scoring system, i.e., the number of AV engines flagging a sample, to select relevant samples to create a dataset. Recent studies~\cite{zhu2020measuring} show that defining a threshold on the score for sample selection is challenging, mainly because the score of the same sample can change unpredictably over time. 
\Cref{fig:vt_score} shows the variation of the VT score after $x$ days following the upload date in four scenarios: (i) the score of malicious Microsoft Word (MS) documents with malware provided by our partners for this analysis; (ii) and (iii) the score of \pPDF s in our dataset, respectively with and without the two largest \cluster s; (iv) the score of PDFs in benign \cluster s.
The data is collected as follows: every day $d_i$, we randomly select up to 500 files per provider--including malicious MS documents--from our dataset up to the day $d_{i-1}$ and submit the selected hashes to VirusTotal to retrieve their VT score. Each file is selected only once.

We observe that documents in the two largest \cluster s, \textit{reCAPTCHA} and \textit{ROBLOX Text}, significantly influence the average score by increasing it to almost twice its value. Without considering these two \cluster s, the overlap between the scores of malicious and benign PDF documents is significant (a histogram of the scores is shown in \Cref{fig:vt_hist1}), making it more challenging to determine an appropriate threshold that could separate them.
Finally, we note that, after 150 days, variance increases, most likely due to the fewer points for older documents.

\subsection{Languages}
\label{sec:targetAnalysis}

We further investigate whether \cluster s target specific geographical areas by using language information obtained via the Google Vision API ~\cite{VisionAI33:online} when processing the first page of each document. We preferred this approach over the extraction of text from the PDF file itself, as the latter approach may lead to incomplete results due to the lack of text in embedded images. Google Vision processed \empirical{\numprint{174298}} images, identifying in total \empirical{62} different languages, \empirical{15} of which with a high confidence threshold  (\empirical{0.90} or higher). Google Vision could not detect text in \empirical{\numprint{678}} documents and could not identify the language in \empirical{131} documents. Results are in \Cref{tab:lang_distribution,tab:global_campaigns,tab:regional_campaigns}.

\begin{table}
	\centering
	\footnotesize
	\begin{minipage}{.6\linewidth}
		\centering
		\begin{tabular}{l r r }
			\multicolumn{3}{c}{\emph{All \cluster s}}\\
			\toprule
			Lang. & Vol. & \# of \cluster s\\
			\midrule
			en       &  \numprint{167475} &    40 \\
			ru       &    1462 &     9 \\
			es       &     757 &    11 \\
			fr       &     211 &     5 \\
			pt-PT    &     154 &     5 \\
			de       &      74 &    10 \\
			it       &      64 &     5 \\
			\bottomrule
		\end{tabular}
		\label{tab:all_campaign_lang}
		\caption*{(a)}
	\end{minipage}%
	\begin{minipage}{.4\linewidth}
		\centering
		\begin{tabular}{l r}
			\multicolumn{2}{c}{\emph{reCAPTCHA only}}\\
			\toprule
			Lang. & Vol.\\
			\midrule
			en       &     \numprint{77425} \\
			es       &       567 \\
			fr       &       165 \\
			pt       &       137 \\
			id       &        96 \\
			it       &        51 \\
						\bottomrule
		\end{tabular}
		\label{tab:recaptcha_distribution}
		\caption*{(b)}
	\end{minipage}
	
	\caption{(a) Distribution of documents per language code (with \# of \cluster s $\ge 5$); (b) Distribution of languages in the \textit{reCAPTCHA} \cluster{} (with \# of documents $\ge 50$).}
	\label{tab:lang_distribution}
\end{table}

We observe that all large-size \cluster s are multi-regional, targeting users in different countries, and that languages are not evenly distributed across documents and \cluster s. English is by far the most common language, covering \empirical{95\%} of the dataset and \empirical{40} \cluster s, followed by Russian (\empirical{0.8\%} and \empirical{nine} \cluster s) and Spanish (\empirical{0.4\%} and \empirical{11} \cluster s).
Small and medium-sized \cluster s tend to focus on \empirical{one} or \empirical{two} languages only (mostly English, \empirical{37} \cluster s, Russian and Spanish, \empirical{eight} \cluster s), except for \textit{CLICK-HERE}, \textit{NSFW `Play Button'} and \textit{Ebooks} which target, respectively, \empirical{17}, \empirical{nine} and \empirical{eight} languages.
When comparing with the distribution of languages on the Internet (see, i.e.,~\cite{stats:mostCommonLangInternet}), we observe that highly-represented Internet languages are virtually not represented in our dataset:  Chinese, the second most used language on the Internet, with about \empirical{19.4\%}, is absent from our malicious documents.

	\section{Distribution Vectors} %
\label{sec:campaignDistributionVectors}

In this section, we present two experiments to confirm the use of two distribution vectors. In $\S$ \ref{sec:email_attach}, we look at the VirusTotal tags of our files, and we search for our file hashes in a corporate spam trap to identify which \cluster s may be distributed as attachments. Then, in $\S$ \ref{sec:seoPoisoning}, we go through search engine results looking for \pPDF s distributed via Search Engine Optimization (SEO) attacks.

\subsection{PDFs as Attachments}
\label{sec:email_attach}

\paragraph{Methodology.} 
The ideal means to determine if our \pPDF s are attached to phishing emails is by using large phishing email datasets, e.g., the Gmail dataset used by Simoiu et al.~\cite{simoiu2020targeted}, which is hard to get in practice, or subscribing to services specialized in malicious email feeds, e.g., MX Mail Data~\cite{mxmaildata}, which costs tens of thousands of dollars. 

As email phishing campaigns target a large number of addresses at once~\cite{simoiu2020targeted}, we speculate that spam traps might also contain phishing emails with attachments. Based on this observation, we asked \PartnerA{} to search for our file hashes inside their spam traps. Also, a closer look at the VT Public API reveals that VT users can upload samples and use the \texttt{attachment} and \texttt{email-spam} tags to indicate the source of the sample~\cite{Filesear14:online}. Accordingly, we use VT tags as an additional data source in this analysis.

 \begin{table}
	\centering
	\footnotesize{
		\begin{tabular}{l rrr}
			\toprule
			{} & 
			\multicolumn{2}{c}{Spamtrap} & 
			\multicolumn{1}{c}{\texttt{attachment}} \\
			{} &
			\# hits & 
			\# PDFs &
			\# PDFs \\
			\midrule	
			AS PDF / File \#1             &               8 &                  2 &       0 \\
			Shared Excel                  &               3 &                  1 &       0 \\
			Amazon scam                   &              15 &                  5 &       0 \\
			Apple receipts                &               5 &                  4 &       0 \\
			PDF Blurred                   &               8 &                  4 &       10 \\
			Fake SE                       &              16 &                  1 &       0 \\
			NSFW `Find'                   &               6 &                  6 &       2 \\
			NSFW `Play'                   &              42 &                 31 &       9 \\
			Try Your Luck                 &               1 &                  1 &      22 \\
			NSFW `Click'                  &               0 &                  0 &      18 \\
			Web Notification              &               0 &                  0 &   2 \\
			\bottomrule
		\end{tabular}
	}
	\caption{\Cluster s with at least two documents marked as \texttt{attachment} or found in a spamtrap by \PartnerA{}.}
	\label{tab:vt_att}
\end{table}

\paragraph{Results.} \Cref{tab:vt_att} shows the result of our experiments.
The total number of matches in \PartnerA{}'s spam trap is \empirical{\numprint{106}} for \empirical{57} unique PDF files, covering \empirical{11} \cluster s. Using a more conservative threshold of at least two matches per file, we have \empirical{68} matches for \empirical{19} files, covering \empirical{seven} \cluster s.
Next, we look at VT tags and use the same data we collected in $\S$ \ref{sec:virustotal_score}, i.e., \empirical{\numprint{106062}} files (\empirical{60.19\%} of our dataset). In total, we found \empirical{\numprint{65}} files with the \texttt{attachment} tag and no files with the \texttt{email-spam} tag, covering \empirical{eight} \cluster s.
Using the same conservative threshold (of two matches) as in the previous analysis, we count \empirical{six} different \cluster s.
Overall, our analysis identified \empirical{11} \cluster s where at least one of the two methods identified at least two PDF files as attachments. \empirical{Two} of these \empirical{11} \cluster s are identified by both methods.

\subsection{SEO Attacks}
\label{sec:seoPoisoning}

A closer look at the PDF documents of the three largest \cluster s (i.e., \textit{reCAPTCHA}, \textit{ROBLOX Text} and \textit{ROBLOX Picture}, covering about \seoCampaingPercentage{}\% of our dataset) reveals that they share distinguishing characteristics with SEO attacks. The first characteristic is \emph{keyword stuffing}~\cite{webspam}, where the resource content is filled with keywords that are relevant to popular searches, ranking the page higher within search results for the included terms. %
We also observe that our PDF files use keywords that are related to the document titles. For example, the keywords used in a document with the title \texttt{Windows xp iso 32 bit file download} can be  \texttt{Microsoft}, \texttt{ISO\_Windows\_XP\_SP3}, and \texttt{crack}.
The second characteristic is \emph{cross-linking resources}~\cite{linkfarm}, which exploits the link-based ranking algorithms of search engines. Attackers craft a network of ad-hoc resources and cross-link them to influence the ranking of target resources. A manual inspection of a sample of documents of the three main \cluster s revealed a consolidated structure of these PDFs, where the first page usually embeds one \textit{bait} link, while the following pages include a list of URLs pointing to other PDFs of the same \cluster{}.
The third characteristic is the \emph{use of benign websites} to host the cross-linked resources~\cite{deseo}, as search engines tend to rank them more quickly than newly registered domains. We verified via GSB~\cite{SafeBrow41:online} that the URLs to these PDFs and the hosting website are not flagged as malicious.

Based on these three observations, we hypothesize that the three largest \cluster s are distributed via SEO attacks and perform a number of experiments to confirm our hypothesis.
We verified that document types that are typically utilized in phishing attacks to infect victims' machines (e.g.,~\cite{le2014look, le2017broad}) do not present the same SEO-oriented document structure by inspecting \empirical{225} MS Word, Excel and OLE2 documents, provided by \PartnerA{}.
We first present the methodology we followed and then our findings.

\paragraph{Methodology.} 
The goal of our experiments is to verify if victims can find \pPDF s belonging to the three largest \cluster s in our dataset via search queries on popular search engines. 
We use as search query the exact string of the document title since we aim at finding direct matches with the \pPDF s in our dataset.
A challenge to the formulation of appropriate search queries is the popularity of the search terms.
Search terms for poisoned search results usually have a lifespan of at most five days, with few exceptions (median: 19 days)~\cite{4year}.
Because titles extracted from VT \pPDF s might not be popular search terms anymore, or the PDFs corresponding to those search queries might have been taken down, we create effective queries with the title of fresh \pPDF s.
The freshness property is ensured through daily selection of newly-uploaded \pPDF s from a new source, i.e., large PDF directories, which we discover by inspecting URLs in \pPDF s in the VirusTotal feed. Specifically, we observe that the URLs in \pPDF s in pages after the first, in the three largest \cluster s,  point to \texttt{.pdf} files.
Many of these URLs share the domain and path, suggesting the existence of large directories hosting cross-linked PDFs. We identify the precise URL of the directory starting from a link pointing to a \texttt{.pdf} file by, first, removing the file name and then, gradually, by removing URL path segments.
This procedure identified \possibleDirectories{} potential URLs of open directories. We verify that the directory index page exists and, if so, that it lists other PDF files hosted on the same directory. 
Then, we ensure that these PDFs are actually \pPDF s. We download each newly uploaded PDF and check if it contains a similar cross-link structure, i.e., if it contains at least \minUrlHeuristic{} URLs, where \empirical{10} end with \texttt{.pdf} but the first one does not. The reason for this threshold is to include as many documents as possible (the average number of URLs ranges from \avgURLrecaptcha{} for \emph{reCAPTCHA} to \avgURLroblox{} for \emph{ROBLOX Picture}). If the PDF file matches our criteria, we extract the title string by parsing the PDF structure. Appendix \ref{appendix:SE-queries} provides additional details on our query search terms.

We monitor index pages daily recording new uploads of PDF files, observing a total of \totalPdfCrawled{} PDF files from Dec. 1st, 2021 to Jan. 30th, 2022. In total, we found \foundIndexPages{} index pages online during the whole duration of the analysis, with a few exceptional downtimes of 1-2 days. However, only \dirWithUpload{} of them had new files uploaded during our study period.
We point out that \emph{we do not store any new PDF files on disk}. Instead, we perform the entire analysis in memory to minimize the risk of fetching documents that are not part of the three targeted \cluster s. We manually verified the accuracy of our heuristic by inspecting a daily sample of ten URLs to determine if the corresponding PDF files belong to the three \cluster s. We conclude that our heuristic is accurate and that all files belong to one of the three \cluster s.

Finally, we use the title string to search for PDF files via web APIs of search engines. In this experiment, we used the web APIs of the two most popular search engines, Google and Bing~\cite{SearchEn9:online}. Each query returns the first top ten results, which we analyze in two ways to determine if an entry contains a PDF file belonging to one of the three \cluster s. First, we check if the result set contains the exact URL of the PDF file. Second, we download the PDF files, checking if they meet the cross-link structure criteria. 

\begin{table}
	\centering
	\footnotesize
	\begin{tabular}{llrr} 
		\toprule
		\makecell[l]{Search engine} & \makecell[l]{Type of match}                                                               & \makecell[r]{Total} & \makecell[r]{Daily Avg}  \\ 
		\hline
		\makecell[l]{Google}        & \makecell[l]{\multirow{2}{*}{Exact match}}                                                 & \makecell[r]{\exactGoogle{}}     & \makecell[r]{\avgExactGoogle{}}       \\ 
		\makecell[l]{Bing}          &                                                                              & \makecell[r]{\exactBing{}} & \makecell[r]{\avgExactBing{}}      \\ 
		\hline
		\makecell[l]{Google}        & \makecell[l]{\multirow{2}{*}{\begin{tabular}[c]{@{}l@{}}Cross-link heuristic\end{tabular}}} & \makecell[r]{\heuristicGoogle{}}   & \makecell[r]{\avgHeuristicGoogle{}}      \\ 
		\makecell[l]{Bing}          &                                                                              & \makecell[r]{\heuristicBing{}} & \makecell[r]{\avgHeuristicBing{}}     \\
		\bottomrule
	\end{tabular}
	
	\caption{Search engines results.}
	\label{tab:search_engines_results}
\end{table}

\begin{table*}
	\centering
	\footnotesize
	\begin{tabular}{c c}
		\includegraphics[width=0.48\linewidth]{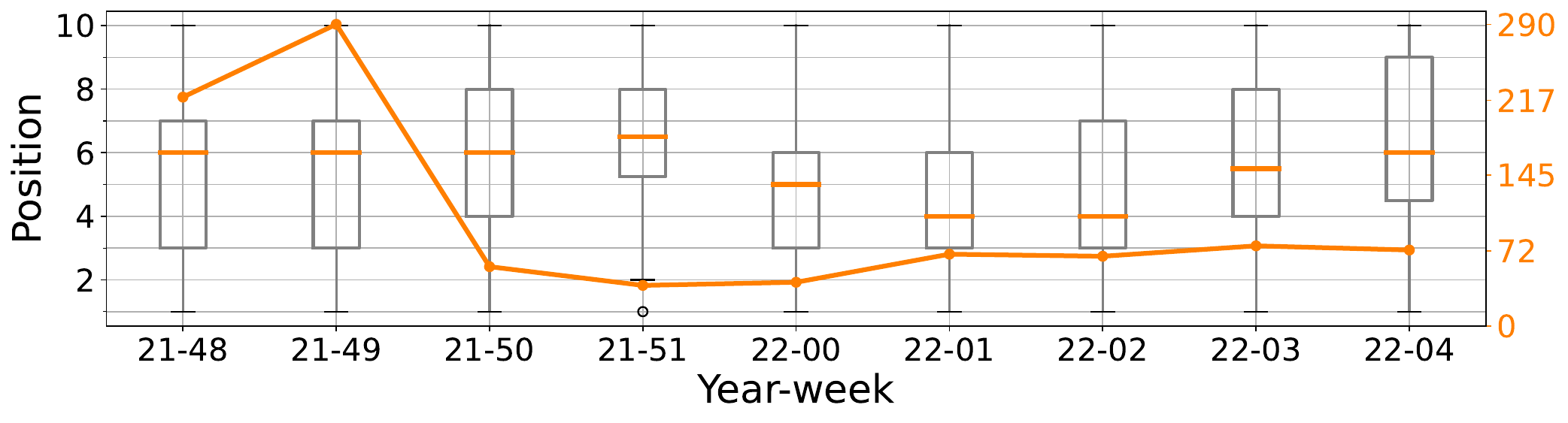}& 
		
		\includegraphics[width=0.48\linewidth]{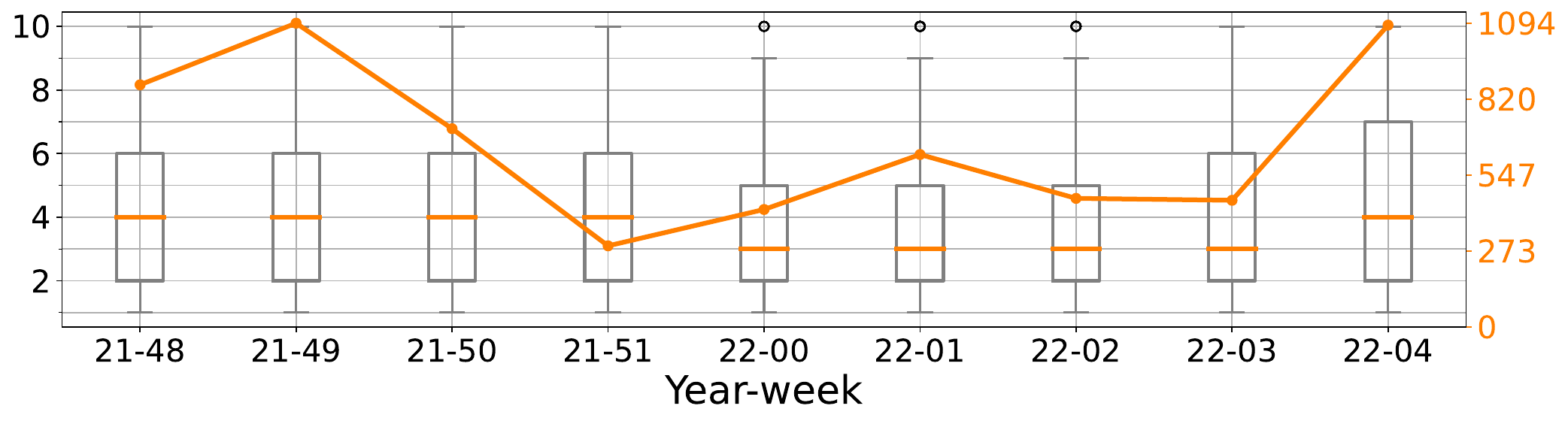}\\
	\end{tabular}
	\captionof{figure}{Number and position of PDFs found on Google (on the left) and Bing (on the right) over time.}
	\label{tab:box-plot}
\end{table*}

\paragraph{Results.} 

\Cref{tab:search_engines_results} shows the number of matches obtained either by exact URL match or by examining the cross-link structure of PDFs. %
In total, we submitted \totalQueries{} queries to each search engine, with differing results depending on the matching heuristic. In total, we successfully retrieved \exactTotal{} documents via exact URL match and \heuristcTotal{} via cross-link heuristic match, confirming our hypothesis that SEO attacks are used in practice. However, results vary across search engines. In general, we observe that finding these PDFs via Google search queries is more challenging than via Bing. In particular, we were not able to retrieve documents on Google via exact URL match, but only via the cross-link heuristic. 

After confirming our hypothesis, we measure the effectiveness of SEO attacks, looking at the ranking of the query results. \Cref{tab:box-plot} shows the weekly number of newly discovered PDFs and their result rank as a box plot. Overall, almost all \pPDF s are ranked high in the query results. Also, we notice a different behavior of Bing and Google, where the average position of PDF files is more stable and higher for Bing than for Google.

	\section{Discussion}
\label{sec:discussion}

This study presents the first categorization of \pPDF s, including an analysis of their distribution vectors.
In this section, we summarize our main findings, evaluate existing defenses, and discuss how to move forward.
Finally, Appendix \ref{app:limitations} further discusses possible limitations of our study.

\subsection{Main Findings}

The main finding of our study is providing sufficient evidence that \pPDF s are not just simple tools within phishing email campaigns. In fact, among \pPDF s, we discovered three \cluster s with unique features in terms of size, duration, and distribution means, indicating the rise of a new kind of web-based \pPDF{} attacks. Below, we present our main results.

\paragraph{\textbf{Many Well-defined \PPDF{} \Cluster s.}} Our study identifies \empirical{\nMaliciousCampaigns{}} \pPDF{} \cluster s, covering nearly all documents in our dataset: \empirical{97\%} of the documents are part of a malicious \cluster{}. Most of the \cluster s are small, with few notable exceptions, e.g., \emph{reCAPTCHA}, \emph{ROBLOX Text} and \emph{ROBLOX Picture}, with 78k, 59k, and 18k files, respectively. Also, we found that large \cluster s tend to be more persistent, with daily uploads. 

\textit{\textbf{More \Cluster s Than Previous Study.}} When comparing our results to the Unit 42 blog post~\cite{paloalto_pdftrend}, our study found \empirical{39} additional \cluster s, including two large ones, i.e., \textit{ROBLOX Text}, and \textit{ROBLOX Picture}.

\textit{\textbf{The Distribution Vector: SEO Attacks.}} Our study confirms that, just as for MalPDF, \pPDF s can be distributed as attachments, by finding files of \empirical{16} \cluster s in a corporate spam trap or flagged as malicious attachments by VirusTotal. However, our study also shows that the three largest \cluster s (i.e., \emph{reCAPTCHA}, \emph{ROBLOX Text}, and \emph{ROBLOX Picture}), covering \empirical{\seoCampaingPercentage{}\%} of our dataset, are distributed via SEO attacks. As we observed, these attacks rely on cross-linked PDF files, requiring the generation of many files for the attack to be effective and explaining the large imbalance of sizes between the top three \cluster s and the others.

\textit{\textbf{\PPDF s Exploit the Web Context.}} \empirical{Ten} \cluster s include UI controls and visual signals commonly observed in webpages, e.g., reCAPTCHA, Google Drive search bar, threaded forum discussions, online repositories for files and torrents, and Web video players. The use of these elements suggests that attackers may expect victims to visualize these documents inside a browser, tricking them into interacting with these elements as with normal web pages.

\subsection{Existing Defenses and Future Directions}
We observed that \pPDF s distributed by SEO attacks represent a persistent threat for victim users.
In this section, we consider existing in-browser defenses (i.e., blocklists) and evaluate the level of protection they offer against attacks delivered by \pPDF s.
Our inspection shows that blocklists offer partial protection against \pPDF s, both in terms of the observed attacks and of the URLs known to the blocklist. We discuss possible roadblocks and future directions for research.

\paragraph{URL Blocklists.}
A quick evaluation of two popular protection systems, Google SafeBrowsing~\cite{SafeBrow41:online} and the rule-based ad-blocking provided by EasyList and EasyPrivacy~\cite{easylist} shows that blocklists offer partial protection against attacks conducted via \pPDF s, with a higher success for websites with malicious advertisements.

Google SafeBrowsing offers a lookup API returning the current blocklist status for a URL and does not provide historical records. However, VirusTotal includes GSB records of the last URL scan in its reports. We observed a low number of matches by using the reports fetched in $\S$ \ref{sec:url_valid}, where \empirical{155} of \empirical{868} URLs (18\%) were blocklisted by GSB,  with \empirical{22} labeled as \textit{malicious} and \empirical{133} as \textit{phishing}.

Ad-block based blocklists provide an additional defense to users by blocking requests to resources matching URLs or patterns in the blocklist.
We logged all outgoing requests when loading the page as we manually inspected websites in $\S$ \ref{sec:url_valid}. Then, we retrieved EasyList and EasyPrivacy blocklists via the Wayback Machine~\cite{internetarchive}, considering the closest available day to the processing date of the PDF file. By matching the collected URLs to the blocklists, we observed that 40\% of the malicious URLs had at least one blocked request. 
These URLs mostly deliver malicious ads or lead to adult sites. 
We further inspected the impact, in terms of potential breakage, of blocked background requests and observed that \empirical{50\%} of these websites were affected, either not loading or stripped of their advertisements. While effective against malicious advertisement and data harvesting sites, ad-blockers fail to protect users against other attacks delivered by \pPDF s.

\paragraph{PDF Detection via Structural Features.}
We also evaluated the effectiveness of existing open-source state-of-the-art malicious PDF detectors~\cite{vsrndic2013detection, chen2020training} in our context.
Established techniques~\cite{vsrndic2013detection, smutz2012malicious} leverage the identification of groups of PDF objects (or ``subtrees'') that are common among malicious PDFs but absent in benign files, often embedding malicious code such as exploits or JavaScript.
Recent advancements~\cite{chen2020training} offer flexibility in this similarity metric, allowing variations such as $N$ differing PDF objects. 

We evaluated Hidost's~\cite{vsrndic2013detection} ability in detecting malicious PDFs or identifying structural similarities among PDFs in the same \cluster{}.
We manually inspected graphical representations and raw PDF objects of sampled files, observing differences in the number, type, and connections of PDF objects across samples, despite visual similarities.
Our analysis of the subtrees identified by the feature selection procedure revealed that they encode specific rendering instructions or metadata objects, which we deem to be a byproduct of the specific PDF generation tool. The feature selection algorithm likely did not identify representative subtrees encoding malicious functionality as MalPDFs are a negligible fraction of our dataset (see $\S$ \ref{sec:dataset}).
The detection result seemed to only loosely correlate with both features of the attack, i.e., the URL leading to malicious activity and the visual bait. This was evident in two ways: first, we could craft proof-of-concept \pPDF s with known URLs and identical visual bait that remained undetected. Second, it successfully identified shared subtrees in PDFs with different visual baits generated with the same tool.
The improvements presented in~\cite{chen2020training} did not lead to better results, as they concern the similarity metric and not the feature selection.
In conclusion, although existing methods such as~\cite{vsrndic2013detection, chen2020training} effectively group PDFs based on structural similarities, they are not suited to our context, as the features of PDF structures lack the necessary discriminatory power to distinguish between benign and \pPDF s, or effectively differentiate \pPDF s belonging to different \cluster s.

\paragraph{Domain-Specific Detection Features.}
Our insights show that existing detection methods for MalPDFs are sub-optimal (see  above), and also that existing commercial solutions lag behind (see $\S$ \ref{sec:virustotal_score} and above).
Nonetheless, our study highlights other distinctive features of \pPDF s that could be integrated into existing detection systems. For example, the three largest categories all include, in pages after the first, a large number of URLs pointing to similar \pPDF{} files, hosted on benign websites (see $\S$ \ref{sec:seoPoisoning}).
One solution could be the joint use of multiple indicators, such as the presence of cross-linked PDFs when they also exhibit visual similarity to known \pPDF{} \cluster s. This information could be used by, e.g., anti-phishing entities or search engines to either maliciously flag or reduce the rank of \pPDF s distributed via SEO attacks. A lower rank in search results could help reduce the number of victim users exposed to \pPDF s as result of queries containing poisoned search terms.

\paragraph{Coverage.}
\label{sec:discussion_VT_coverage}

Our findings in $\S$ \ref{sec:seoPoisoning} show the result of an ongoing malicious activity, where \pPDF s can be found on popular search engines when querying for specific popular keywords. We thus investigated if those PDFs had already been discovered by an anti-phishing entity and uploaded on VirusTotal by looking for \texttt{SHA256} matches between the \pPDF s found on search engines (\numprint{3112} files) and those in our dataset.
A total of \empirical{44} PDFs were already known to VT among those found on search engines, \empirical{17} of which were known to VT from \empirical{10} days to \empirical{eight} months prior.
These empirical observations are in line with the findings presented in $\S$ \ref{sec:duration_and_activity}, i.e., the activity of most \cluster s lasts for a long time, even extended to the online availability of single PDF files. 
Conversely, \empirical{27} PDFs observed in our search results later appeared in our partners' feeds, with an average delay of \empirical{22} days.
The reasons for the limited overlap may lie in different concurrent causes, e.g., the PDFs were not flagged as malicious on their first submission or did not receive a `phishing' label (a criterion of \PartnerB{}). Alternatively, they may have been uploaded after the end of our data collection period.

Nonetheless, the crowdsourced nature of VT and the filtering rules employed by our partners may have introduced a source of bias in our data collection. We believe this bias may be evident in the amount of data, i.e., the size of this phenomenon may be bigger than our measurements report. Conversely, independent studies, like the one of Palo Alto Networks~\cite{paloalto_pdftrend}, report results similar to ours in terms of discovered \cluster s, which corroborates our findings.

\paragraph{Future Directions.}
Overall, we observed that the coverage of the phenomenon of \pPDF s is not exhaustive. This may be due to the combined medium of PDF binary and web page delivering the attack, and to the diverse nature of the attacks \pPDF s lead to.
The low coverage of the inspected URL blocklists may be due to their incompleteness, given by the inability of ecosystem players to extract URLs from PDF files and feed them back to blocklists.
In fact, the few URLs flagged as malicious (by GSB or VT) may be attributed to manual submissions.
This shortcoming may result from the good reputation held by hosting providers, which can make blocklisting challenging. %
Nonetheless, a closer look at the autonomous system names hosting the 868 URLs flagged as malicious suggests the opposite, as they include popular providers such as Cloudflare, AWS, and Google Cloud Platform. This conflicting observation reaffirms the need for more research in this field to determine the role, reach and limitations of anti-phishing ecosystem players.

\subsection{Data Sharing and Ethics}
\label{subsec:partners_agreement}

Two industrial partners provided the samples of our dataset. While we are not allowed to share the raw PDF files, we can publish the metadata of our dataset allowing researchers to reproduce and build on our results. We will share all file hashes of the PDF files (allowing to retrieve them from VirusTotal), PDF file screenshots, clustering labels and URLs. The data and supporting scripts can be found at \url{https://www.kaggle.com/datasets/emerald101/from-attachments-to-seo} .

This study did not involve human subjects, and we did not seek IRB involvement. However, we discuss a few ethical considerations of our study.
One concern of our study is that VirusTotal files may contain private data. While VT allows the removal of private files, there is a possibility that they ended up in our dataset. Our manual evaluations exclude that \pPDF s (\empirical{98.94\%} of the files) contain private information; still, the non-malicious ones might contain such information. Before releasing the dataset, we will manually inspect the remaining \empirical{\numprint{1862}} benign PDFs, removing those with private information.

Another concern is that the SEO attack experiments may have downloaded files with private information. We addressed this concern at the design time, enforcing two strict rules: (i) we process PDF files only in memory, and (ii) we use our cross-link heuristic to guarantee that we store the metadata, e.g., URLs and file hash, only of those files fitting the heuristic. %
Finally, we retrieved contact points for those websites hosting direct \pPDF{} matches (observed in $\S$ \ref{sec:seoPoisoning}) and raised awareness of the ongoing threat following the state of the art for vulnerability notifications~\cite{stock2018didn, li2016you}.

	\section{Related Works} %
\label{sec:related_work}

We now review works closely related to our study. 

\paragraph{\PPDF s and MalPDFs.} The closest study to ours is the non-peer-reviewed analysis~\cite{paloalto_pdftrend} performed by Unit 42 of Palo Alto Networks. Unit 42's analysis indicated a surge of \pPDF{} files, illustrating the existence of five \cluster s and analyzing the landing pages of the \emph{reCAPTCHA} one. In comparison, our study relies on a dataset that is dwarfed by the 
\empirical{5.2} million files of Unit 42 (about \empirical{0.033\%}). Nevertheless, we not only confirm the presence of the five \cluster s but discover \empirical{39} new malicious ones, including two large \cluster s, \emph{ROBLOX Text} and \emph{ROBLOX Picture}, which were not found by Unit 42. In addition, our results help build a better picture of the \pPDF{} ecosystem, showing that VirusTotal scores are of little help, as opposed to scores for malicious Office documents. Last but not least, our study tackles the question of distribution, confirming the use of attachments and showing the use of another distribution vector, i.e., SEO attacks.

The analysis of MalPDFs is also a research area close to our work. Several works (e.g.,\cite{maiorca2012pattern, vsrndic2013detection, smutz2012malicious}) proposed MalPDFs detection via machine learning, leveraging features derived from the internal structure of PDF documents, or relying on the analysis of embedded JavaScript~\cite{laskov2011static, carmony2016extract, tzermias2011combining}. Other works show how to evade existing classifiers (e.g.,~\cite{xu2016automatically, carmony2016extract, laskov2014practical}) or how to improve their robustness (e.g.,~\cite{chen2020training, smutz2016tree}). M\"{u}ller et al.~\cite{muller2021processing} crafted MalPDFs exploiting caveats in the PDF specification, also without using JavaScript or exploit code. In our work, we do not focus on this type of documents, and we estimate their presence in our dataset to be very low.

\paragraph{Phishing Attacks.} %
Many works tackled the detection of phishing messages and web pages, i.e., detection of malicious emails (e.g.,~\cite{fette2007learning, ho2017detecting, cidon2019high, khonji2012enhancing, duman2016emailprofiler}), URL and page content (e.g.,~\cite{whittaker2010large, liang2016cracking, xiang2011cantina+, zhang2007cantina}), and passive DNS data (e.g.,~\cite{bilge2011exposure}). Recently, new ideas proposed using visual features of web pages to find phishing attacks (e.g.,~\cite{abdelnabi2020visualphishnet, lin2021phishpedia, liu2022inferring}). %
As opposed to these works, our paper does not present a technique to detect phishing attacks nor does it evaluate anti-phishing techniques. Our paper provides the first characterization of the threat posed by \pPDF{} files. 
Other studies focused on characterizing victims of phishing emails~\cite{simoiu2020targeted} and why they fall for phishing~\cite{blythe2011f}, on measuring the effectiveness of such campaigns~\cite{oest2020sunrise}, on visual features of malicious links in social networks (e.g.,~\cite{stivala2020deceptive}), on victims' characteristics on social networks, such as gender, age, and country (e.g.,~\cite{redmiles2018examining}), and cognitive response to malicious emails (e.g.,~\cite{van2019cognitive}). %

\paragraph{Email Phishing Campaigns.}
Simoiu et al.~\cite{simoiu2020targeted} study phishing campaigns delivered by email and measure their volume and duration. In this respect, our works are related. 
The \pPDF{} \cluster s of this paper show significant differences compared with email-based campaigns in terms of volume, duration, and temporal features. \PPDF{} \cluster s are lower in number and larger in size; they last longer and they do not happen in bursts, but they are rather constant or show less frequent and well-distanced peaks. %

\paragraph{SEO Attacks.} SEO attacks are used to game page rankings to expose users to a variety of Web attacks and campaigns (see, e.g., ~\cite{deseo,4year,seobotnet,scalable_detection}). Our work shares similarities with, e.g.,~\cite{pbn,linkfarm, webspam}, as \pPDF s also employ document cross-linking or keyword stuffing to game SE ranks. Even if an update in the Google Pagerank algorithm made keyword stuffing-based attacks largely ineffective~\cite{fashion_crimes}, Liao et al.~\cite{longtail} demonstrated how malicious players can still find new ways into search results. 
	
	\section{Conclusion} %
In this paper, we presented the first comprehensive study and categorization of \pPDF s, quantifying the threat posed by this kind of malicious PDF documents.
We identified \nMaliciousCampaigns{} \pPDF{} \cluster s in a real-world dataset of \nAnalyzedSamples{} PDFs and studied their volumetric and temporal properties.
We observed large-size, long-lasting \cluster s, active for almost the entire duration of our study, and highlighted their difference with respect to email phishing \cluster s.
Further, we studied the visual baits in \pPDF s and observed that several \cluster s include visual elements typical of web pages, e.g., fake reCAPTCHA buttons. In addition, we assessed the usefulness of online scoring systems such as the one provided by VirusTotal.
Finally, we performed a series of experiments studying the distribution vectors used by attackers.
Overall, our main finding consists in providing enough evidence that \pPDF s mainly spread through SEO attacks (\seoCampaingPercentage{}\% of our dataset), while we observe their usage as part of email campaigns on a much lower scale.
We publicly release the screenshot dataset, metadata and labeling performed during this study to foster new research on this subject.

	\section*{Acknowledgments} We thank Pedram Amini (InQuest) for providing the data of this study, and thank the InQuest team for their valuable feedback.

	\bibliographystyle{ACM-Reference-Format}
	\bibliography{bibliography} 
	
	\appendix
	\renewcommand{\thesubsection}{\Alph{subsection}}
	
\begin{figure*}
	\centering
	\includegraphics[width=0.9\textwidth]{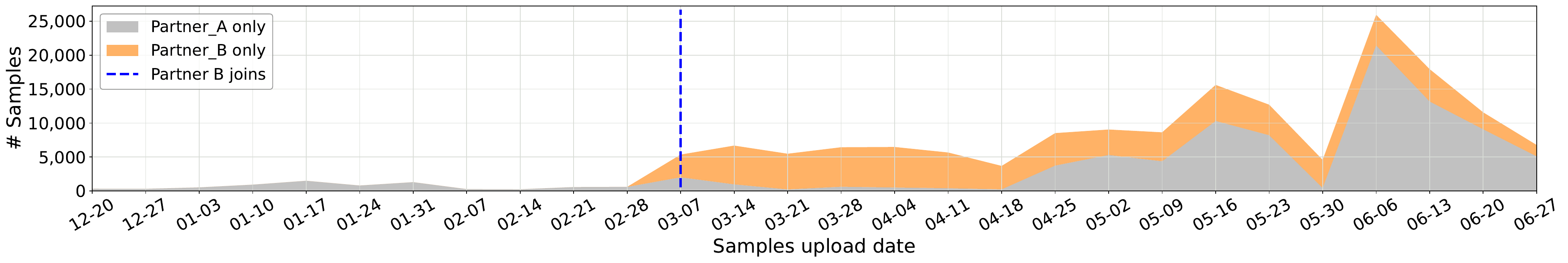}
	\caption{Weekly sum of daily uploads of the two datasets, stacked.}
	\label{fig:samples-over-time}
\end{figure*}

\subsection{PDF Clustering}
\label{sec:clustering_details}

In this section, we expand on the procedure used to cluster visually similar documents, described in $\S$ \ref{sec:campagin_identification}.
First, we report the evaluation on the embeddings returned by DeepCluster~\cite{caron2018deep}. Then, we explain which parameters were used to run DBSCAN~\cite{ester1996density} and why. Finally, we report on our validation of the obtained clusters.

\begin{figure*}
	\centering
	\includegraphics[width=0.9\textwidth]{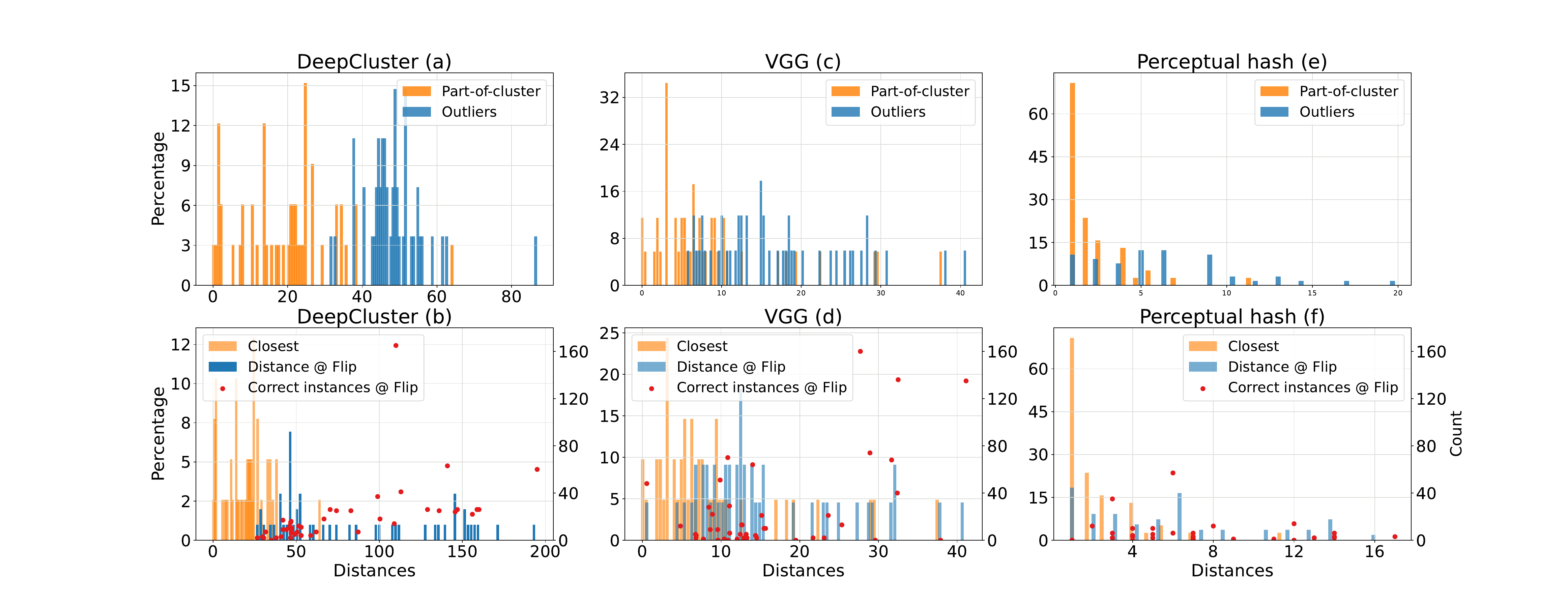}
	\caption{Comparing: (above) closest distances for outliers and `part-of-\cluster{}' samples, (below) closest distances of `intra-\cluster{}' samples and distances at \cluster{}-flipping points. Scatter plots display correct instances (same \cluster{}) at flipping points.}
	\label{fig:metric-distances}
\end{figure*}

\paragraph*{DeepCluster Embeddings Evaluation.} 
As the dataset does not have ground-truth labels, we evaluate the embeddings by visually inspecting the nearest neighbours (using $L_2$ distances) for a small subset; the top nearest neighbours for a document should ideally be visually similar. %
Our evaluation further covers three criteria: (i) Outliers (documents which lack any similarity to others) should have larger closest distances compared to documents that have similar counterparts (intra-\cluster{}). (ii) There should be a consistent distance threshold beyond which the samples are no longer similar to the query (a ``\cluster{}-flipping'' point). (iii) The number of similar images returned before the ``flipping'' point should be as high as possible. We use the results of the $k$-means clustering to inspect and select representative samples from different visual \cluster s (51 samples) and from outliers (48 samples of documents that were not grouped with similar ones). We compare three methods that can be used as a metric: the trained DeepCluster network, pre-trained VGG~\cite{simonyan2014very}, and perceptual hashing. 
The first row in \Cref{fig:metric-distances} shows the comparison between outliers and intra-\cluster{} documents. As observed, the trained DeepCluster has a better separation between the two cases, followed by VGG, while perceptual hashing has a very poor one. For intra-cluster documents, the second row shows the distance thresholds at which the \cluster s flipped, in comparison with the smallest distances. Ideally, there should be enough separation on average between the flipping points and the closest distances in order to select thresholds that allow the formation of clusters. This is, again, better accomplished by DeepCluster rather than by VGG and hardly accomplished by perceptual hashing. These three images also show the count of correct images retrieved at the flipping point: while VGG can retrieve more samples than DeepCluster, the threshold distances for VGG are less consistent, making it less useful for further clustering. On the other hand, the retrieved samples in the case of perceptual hashing are relatively few.  

Overall, this analysis shows that the embeddings obtained by training DeepCluster are more useful in terms of nearest neighbours analysis than perceptual hashing and off-the-shelf pre-trained CNNs. It also gives insights into the possible distance thresholds that could be used in a next clustering step.

\paragraph*{Parameters of DBSCAN}
The first parameter needed to run DBSCAN is representative of the minimum number of samples in a cluster. To be able to capture even very small \cluster s, we select a minimum number of samples per cluster of $3$, leveraging the insights gained during the previous step of manual inspection.
When selecting a distance threshold for DBSCAN, we keep into consideration the insights observed during the clustering procedure (see \empirical{Figure 7(b)}). The last intra-\cluster{} sample is located at a relative distance of 70, although several outliers are already present at this threshold. We keep a conservative approach to reduce the number of outliers as well as to maximize the number of correctly classified instances and choose a distance threshold of $50$. This procedure returned \empirical{120} clusters and \empirical{458} noise points, identifying \empirical{nine} new clusters, however, including \empirical{68\%} non-homogeneous clusters.
We therefore finally used a lower distance threshold, i.e. $35$, where most of the samples in the ``Closest'' group are located. In this case, DBSCAN outputted \empirical{\nDBSCANClusters} clusters and \empirical{\nDBSCANnoise} noise points. We manually validate clusters' coherence by inspecting all samples, obtaining \empirical{\nDBSCANHomCl} homogeneous clusters, for a total of 610 documents. It is important to note that while the overall clustering procedure we followed might involve some tuning steps to select the parameters, it drastically reduces the time to label all samples individually and identify all \cluster s in the dataset.

\begin{table*}
	\centering
	{\footnotesize
		\begin{tabular}{l | r | r r | r | l r | l r}
			\toprule
			\multicolumn{1}{c}{}& 
			\multicolumn{1}{c}{Input} & 
			\multicolumn{2}{c}{Clusters} & 
			\multicolumn{5}{c}{New \cluster s}\\
			
			\multicolumn{1}{c|}{Step}& 
			\multicolumn{1}{c|}{Size}& 
			Clust. No.&  
			Docs. No.& 
			Incr. & 
			Largest Campaing Name& 
			Rank& 
			Smallest  Campaing Name& 
			Rank\\  
			
			\midrule
			SHA256 dedup. & \numprint{185575} &   - &                 - &   0 &                              - &  - & -                    &  -\\
			phash dedup.  & \numprint{176208} &   - &                 - &   0 &                              - &  - & -                    &  -\\
			\midrule
			$k$-means     &  \numprint{20671} & 635 &  \numprint{18557} & +15 & reCAPTCHA                      &  1 & Fake SE              & 35\\
			DBSCAN        &   \numprint{2114} &  87 &               610 & +29 & reCAPTCHA Drive                &  6 & Download File        & 47\\
			Full-manual   &   \numprint{1504} &   - &                 - & +36 & AS PDF / File \#12             & 10 & Shared Excel         & 48\\
			Outliers      &               389 &   - &                 - &   0 &                              - &  - & -                    &  -\\
			\bottomrule
		\end{tabular}
	}
	\captionof{table}{PDF \cluster{} identification: overview of the input/output properties of each step.}
	\label{tab:camp_ident_size}
\end{table*}

\paragraph*{Clusters Validation.}
\label{sec:cl_error}

We estimated the overall clustering effectiveness by selecting at most 20 random PDF files for each of the \nLabels{} \cluster s\footnote{\Cluster s can contain fewer than 20 files.}, collecting \numprint{1071} samples, and checking for labeling errors. The fraction of mislabeled samples is 3.27\%, which, in perspective, is about half of the error in popular datasets, e.g., 6\% of ImageNet\cite{northcutt2021pervasive}.

\subsection{False Positives in Maliciousness Validation}
\label{sec:fp_benign_categories}

In this section, we examine the conflicts that arose when the manual validation procedure did not confirm a `malicious' label in VirusTotal reports.
In particular, we examine those \cluster s where not only no malicious activity was observed, but also the visual content lacked any form of deceit.
These \cluster s are: \textit{Book cover}, \textit{Document Layout}, \textit{Invoice-like}, \textit{AS PDF / File \#12}, \textit{Boletín de Noticias}, \textit{Excel tables}, \textit{Informative Flyer}, \textit{Netcraft}.

\paragraph{No Sign of Malicious Activity.} Four \cluster s, i.e., \textit{AS PDF / File \#12}, \textit{Boletín de Noticias}, \textit{Excel tables}, \textit{Netcraft} show no sign of malicious activity. The first \cluster{} groups PDFs designed for phishing training (e.g., within a company) and specifically crafted to be flagged by AVs. In fact, clicking the link embedded in the PDFs leads to a webpage hosted by the organizing company, which reveals that the document was a test and includes educational content on phishing. The second and third \cluster s include links to security-related resources, as they promote educational material.
Similarly, PDFs in the fourth \cluster{} include rich-text dumps of the URL-scanning tool from the security company Netcraft, reporting on malicious sites. 

\textit{Outliers Flagged as Malicious.} Three \cluster s, i.e., \textit{Book cover}, \textit{Document Layout}, \textit{Informative Flyer} include documents whose URLs  have been correctly flagged by VirusTotal. Upon manual inspection, we verified that the \cluster{} label of these documents is not correct---in other words, they are improperly assigned to these \cluster s and, as such, do not contribute to making the \cluster{} a malicious \cluster{}.

\textit{One URL Flagged as Malicious.} In two cases, i.e., \textit{Invoice-like} and \textit{Document Layout}, one URL per \cluster{} was flagged. Upon manual inspection, the URL in \textit{Document Layout} appeared to be flagged by Google SafeBrowsing, while the URL in \textit{Invoice-like} pointed to the main page of a hosting provider. 
We speculate the reason for this may be that these \cluster s aggregate a few documents with larger intra-cluster distances, which alternatively could have been split in sub-\cluster s or moved to the \textit{Outliers} cluster, as the manual validation procedure did not raise any flag.

\begin{table}
	\centering
	\footnotesize{
		\begin{tabular}{l rr|  rr}
			\toprule
			{}                          & \multicolumn{2}{c}{URL Coverage} &  \multicolumn{2}{c}{Document Coverage}\\
			\midrule
			AS PDF / File \#1           &    286 &             100.00\% &   285 &         99.65\% \\
			Book cover                  &    252 &              94.59\% &   248 &         98.41\% \\
			Document Layout             &    322 &             100.00\% &   320 &         99.38\% \\
			Download File               &      3 &              66.67\% &     2 &         66.67\% \\
			PDF Blurred                 &    274 &             100.00\% &   273 &         99.64\% \\
			Ebooks                      &    789 &              97.98\% &   765 &         96.96\% \\
			NSFW `Find'                 &    397 &              99.69\% &   396 &         99.75\% \\
			NSFW `Play'                 &   \numprint{9783} &   49.37\% & \numprint{4827} & 49.34\% \\
			Netcraft                    &    298 &             100.00\% &   281 &         94.30\% \\
			ROBLOX  Picture             &  \numprint{12497} &   14.04\% & \numprint{1829} & 14.64\% \\
			ROBLOX Text                 &  \numprint{36919} &    9.59\% & \numprint{2120} & 5.74\% \\
			Crawler trap                &   \numprint{4917} &   99.35\% & \numprint{1738} & 35.35\% \\
			reCAPTCHA                   &  \numprint{77988} &    1.28\% & \numprint{1000} & 1.28\% \\
			reCAPTCHA Drive             &   \numprint{1692} &   34.79\% &   589 &         34.81\% \\
			
			\bottomrule
		\end{tabular}		
	}
	\caption{Number of bait URLs submitted to VT and respective number of PDFs. Missing \cluster s have 100\% coverage.}
	\label{tab:vt_coverage}
\end{table}

\subsection{Search Engine Queries}
\label{appendix:SE-queries}
Our queries use \empirical{\numprint{15436}} individual keywords.
The most frequent keywords are English words, and among the top five we have \texttt{pdf} (\empirical{\numprint{9270}}), \texttt{free} (\empirical{\numprint{3732}}), \texttt{guide} (\empirical{\numprint{2233}}), \texttt{template} (\empirical{\numprint{1822}}), and \texttt{manual} (\empirical{\numprint{1740}}). When looking at their effectiveness, \texttt{pdf} is used to find \empirical{\numprint{2036}} new documents, followed by \texttt{answers} (\empirical{\numprint{356}}), \texttt{free} (\empirical{\numprint{332}}), \texttt{guide} (\empirical{\numprint{316}}), and \texttt{movie} (\empirical{\numprint{288}}). We also look at the frequency distribution of query bigrams, with the top five most effective words being \texttt{answer key} (\empirical{\numprint{156}} files), \texttt{pdf free} (\empirical{\numprint{116}} files), \texttt{how to} (\empirical{\numprint{109}} files), \texttt{full movie} (\empirical{\numprint{89}} files) and \texttt{edition pdf} (\empirical{\numprint{80}} files).

\begin{table}
	\centering
	\footnotesize
	\begin{tabular}{l | r r}
		\toprule
		Regional \cluster s & Vol. & Lang.\\
		\midrule
		reCAPTCHA Drive                     &  \numprint{1693} &   en \\
		Download Torrent                    &  \numprint{1120} &   ru \\
		AS PDF / File \#1                   &   134 &   en \\
		Access Online Gen.                  &    55 &   en \\
		Lottery 25th Ann.                   &    43 &   ru \\
		AS PDF / File \#4                   &    41 &   en \\
		Apple receipts                      &    30 &   en \\
		NSFW `Dating'                       &    14 &   en \\
		AS PDF / File \#11                  &    11 &   en \\
		AS PDF / File \#3                   &    11 &   en \\
		\bottomrule
	\end{tabular}
	\caption{\Cluster s targeting one language (Vol. $> 10$ docs).}
	\label[table]{tab:regional_campaigns}
\end{table}

\begin{table}
	\centering
	\footnotesize
	\begin{tabular}{l | r r}
		\toprule
		Multi-regional \cluster s           & Vol. & \# Lang.s \\
		\midrule
		reCAPTCHA                       &  \numprint{78852} &     52 \\
		CLICK-HERE                      &    286 &     17 \\
		NSFW `Play'                     &   \numprint{9126} &      9 \\
		ROBLOX Text                     &  \numprint{59345} &      9 \\
		Ebooks                          &    795 &      8 \\
		ROBLOX Picture                  &  \numprint{18065} &      6 \\
		Download Btn                    &     19 &      5 \\
		PDF Blurred                     &    228 &      3 \\
		AS PDF / File \#8               &      6 &      3 \\
		Play Video                      &     70 &      3 \\
		Download PDF                    &     13 &      3 \\
		Coin Generator                  &    167 &      2 \\
		Amazon scam                     &     14 &      2 \\
		Elon Musk BTC                   &     82 &      2 \\
		Web Notification                &      8 &      2 \\
		Try Your Luck                   &     79 &      2 \\
		Russian Forum                   &    167 &      2 \\
		Fake SE                         &     18 &      2 \\
		NSFW `Click'                    &     44 &      2 \\
		NSFW `Find'                     &    322 &      2 \\
		Sigue Leyendo                   &     10 &      2 \\
		AS PDF / File \#6               &      5 &      2 \\
		\bottomrule
	\end{tabular}
	\caption{Multi-regional \cluster s.}
	\label{tab:global_campaigns}
\end{table}

\begin{figure}
	\centering
	\includegraphics[width=0.70\linewidth]{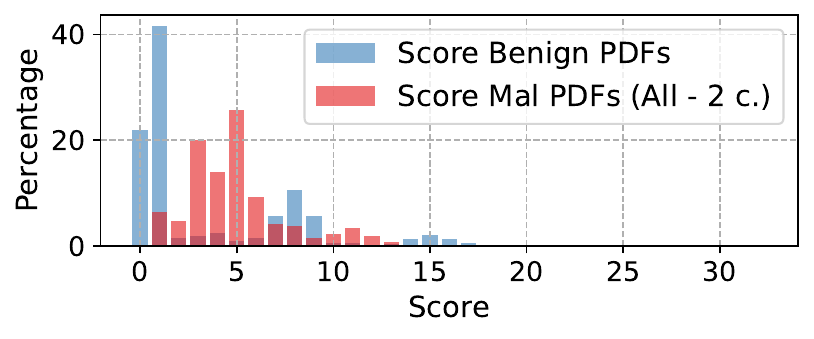}
	\caption{Histograms of the VT scores of benign and \pPDF s, without the two largest \cluster s, for the first 30 days.}
	\label{fig:vt_hist1}
\end{figure}

\subsection{Limitations}
\label{app:limitations}
This study should be considered alongside certain limitations.
Due to an accidental cap limiting the number of PDFs in their feed, \PartnerA{} sent us a maximum of \empirical{300} samples per day, until \empirical{March 3rd}, when this cap was removed. Until then, the dataset accounted for \empirical{\numprint{7787}} unique samples,  i.e., \empirical{4.41\%} of the entire dataset, affecting \empirical{30} of the \cluster s leading to attack pages. 
While this may influence the size of the \cluster s, it did not prevent us from observing \cluster s with samples linking to malicious activity.
Among them, \empirical{four} \cluster s (\textit{Netflix scam}, \textit{Shared Excel}, \textit{Download Btn}, \textit{AS PDF / File \#13}, \textit{Adobe Click}, \textit{AS PDF / File \#10}, \textit{Apple receipts}) saw a contribution of \empirical{50\%} or higher of their entire volume, and, \empirical{three} \cluster s entirely take place before March 3rd.
Moreover, the \textit{reCAPTCHA} \cluster{} has received \empirical{+\numprint{2897}} samples, which is a marginal increase when considering the size of this \cluster{}. Similarly, the \textit{NSFW `Play'} \cluster{} has seen a contribution of \empirical{+\numprint{4262}} samples, corresponding to \empirical{44\%} of its volume.
The remaining \cluster s received a very limited number of samples, on average \empirical{16} samples each.
Nevertheless, including the data points before March 3rd gave us the invaluable opportunity to place the starting date of each \cluster{} at a much earlier point in time (\empirical{45} days on average).%

Before implementing our clustering procedure, we evaluated a series of possibilities.
Using URLs as an additional feature may have improved accuracy, but two main challenges make this inadequate in practice. First, as the same \cluster{} uses different URLs, URL string matching would have resulted in clusters that are too fragmented. Second, using maliciousness scores from online services is also a weak signal for clustering. %
$\S$ \ref{sec:mal_results} shows that URL analysis services are incomplete, making them more suitable for determining the maliciousness of a \cluster{} via random sampling rather than a feature for clustering. 
Finally, visiting all landing pages and detecting attacks requires tackling non-trivial challenges, e.g., bypassing client-side cloaking and detecting malicious pages. 
CrawlPhish~\cite{zhang2021crawlphish}, by Zhang et al., tackles both challenges. Unfortunately, this tool is not available in practice\footnote{The authors could not share the code with us because it relies on a third-party component that they are not authorized to share.}, making it challenging to analyze URLs at scale for our purpose.

\end{document}